\documentstyle[epsfig]{mn}

\newif\ifAMStwofonts

\newcommand{\be}{\begin{equation}}
\newcommand{\ee}{\end{equation}}
\newcommand{\bea}{\begin{eqnarray}}
\newcommand{\eea}{\end{eqnarray}}
\newcommand{\Bea}{\begin{eqnarray*}}
\newcommand{\Eea}{\end{eqnarray*}}
\newcommand{\pa}{\partial}

\newcommand{\na}{\nabla}

\ifoldfss
  \ifCUPmtlplainloaded \else
    \NewTextAlphabet{textbfit} {cmbxti10} {}
    \NewTextAlphabet{textbfss} {cmssbx10} {}
    \NewMathAlphabet{mathbfit} {cmbxti10} {} 
    \NewMathAlphabet{mathbfss} {cmssbx10} {} 
  \fi
  \ifAMStwofonts
    \ifCUPmtlplainloaded \else
      \NewSymbolFont{upmath} {eurm10}
      \NewSymbolFont{AMSa} {msam10}
      \NewMathSymbol{\upi}     {0}{upmath}{19}
      \NewMathSymbol{\umu}     {0}{upmath}{16}
      \NewMathSymbol{\upartial}{0}{upmath}{40}
      \NewMathSymbol{\leqslant}{3}{AMSa}{36}
      \NewMathSymbol{\geqslant}{3}{AMSa}{3E}

      \let\leq=\leqslant 
      \let\geq=\geqslant 
    \fi
  \fi
\fi 

\ifnfssone
  \newmathalphabet{\mathit}
  \addtoversion{normal}{\mathit}{cmr}{m}{it}
  \addtoversion{bold}{\mathit}{cmr}{bx}{it}
  \newmathalphabet{\mathbfit} 
  \addtoversion{normal}{\mathbfit}{cmr}{bx}{it}
  \addtoversion{bold}{\mathbfit}{cmr}{bx}{it}
  \newmathalphabet{\mathbfss} 
  \addtoversion{normal}{\mathbfss}{cmss}{bx}{n}
  \addtoversion{bold}{\mathbfss}{cmss}{bx}{n}
  \ifAMStwofonts
    \ifCUPmtlplainloaded \else
      \UseAMStwoboldmath
      \makeatletter
      \new@mathgroup\upmath@group
      \define@mathgroup\mv@normal\upmath@group{eur}{m}{n}
      \define@mathgroup\mv@bold\upmath@group{eur}{b}{n}
      \edef\UPM{\hexnumber\upmath@group}
      \new@mathgroup\amsa@group
      \define@mathgroup\mv@normal\amsa@group{msa}{m}{n}
      \define@mathgroup\mv@bold\amsa@group{msa}{m}{n}
      \edef\AMSa{\hexnumber\amsa@group}
      \makeatother
      \mathchardef\upi="0\UPM19
      \mathchardef\umu="0\UPM16
      \mathchardef\upartial="0\UPM40
      \mathchardef\leqslant="3\AMSa36
      \mathchardef\geqslant="3\AMSa3E

      \let\leq=\leqslant 
      \let\geq=\geqslant 
    \fi
  \fi
\fi 

\ifnfsstwo
  \DeclareMathAlphabet{\mathbfit}{OT1}{cmr}{bx}{it}
  \SetMathAlphabet\mathbfit{bold}{OT1}{cmr}{bx}{it}
  \DeclareMathAlphabet{\mathbfss}{OT1}{cmss}{bx}{n}
  \SetMathAlphabet\mathbfss{bold}{OT1}{cmss}{bx}{n}
  \ifAMStwofonts
    \ifCUPmtlplainloaded \else
      \DeclareSymbolFont{UPM}{U}{eur}{m}{n}
      \SetSymbolFont{UPM}{bold}{U}{eur}{b}{n}
      \DeclareSymbolFont{AMSa}{U}{msa}{m}{n}
      \DeclareMathSymbol{\upi}{0}{UPM}{"19}
      \DeclareMathSymbol{\umu}{0}{UPM}{"16}
      \DeclareMathSymbol{\upartial}{0}{UPM}{"40}
      \DeclareMathSymbol{\leqslant}{3}{AMSa}{"36}
      \DeclareMathSymbol{\geqslant}{3}{AMSa}{"3E}

      \let\leq=\leqslant 
      \let\geq=\geqslant 
    \fi
  \fi
\fi 

\ifCUPmtlplainloaded \else
  \ifAMStwofonts \else 
    \def\upi{\pi}
    \def\umu{\mu}
    \def\upartial{\partial}
  \fi
\fi
\title[Solar surface simulations]
{3D convection simulations of the outer layers of the Sun using realistic physics.}
\author[Robinson et al.]
       {F. J. Robinson,$^1$\thanks{Email: marjf@astro.yale.edu}
            P. Demarque,$^1$ L. H. Li,$^1$  S. Sofia,$^1$
\newauthor
Y.-C. Kim$^2$, K.L. Chan$^3$ and D.B.Guenther$^4$  \\
$^1$Astronomy Department, Yale University, Box 208101, New Haven, CT 06520-8101
$^2$Yonsei University, Seoul, South Korea\\
$^3$Hong Kong University of Science \& Technology, Hong Kong, China\\
$^4$Department of Astronomy and Physics, Saint Mary's University,
Halifax,
Nova Scotia B3A 4R2, Canada}

\begin{document}
\maketitle

\label{firstpage}

\begin{abstract}
  
This paper describes a series of 3D simulations of shallow 
inefficient convection in the outer layers of the Sun. 
The computational domain is  a closed 
box containing the convection-radiation transition layer, located at the 
top of the solar convection zone.
The most salient features of the simulations  are that: 
i)The position of the lower boundary can have a major effect on the characteristics
of solar surface convection 
(thermal structure, kinetic energy and  turbulent pressure). 
ii)The width of the box
has only 
a minor effect on the thermal structure, but a more significant effect on the dynamics (rms velocities).
iii)Between the surface and a depth of 1 Mm, even though the density and pressure 
increase by an order of magnitude, the  vertical correlation length of vertical velocity 
is always close to 600 km.
iv)  In this region the vertical velocity cannot be scaled by the pressure 
or the density scale height. This casts doubt on the applicability of the mixing length theory, not only 
in the superadiabatic layer, but also in the adjacent underlying layers.
v) The final statistically steady state  is not strictly dependent on the 
initial atmospheric stratification.

\end{abstract}

\section{Introduction}

It is now just possible to perform physically realistic 3D simulations of the 
surface layers of the Sun which take into account the complex 
interaction between radiative and convective energy transports 
(Stein \& Nordlund 1998, Kim \& Chan 1998  hereafter denoted SN and KC,
respectively, Stein \& Nordlund 2000). 
There have also been a number of  2D simulations of the surface layers (Steffen et
al. 1990  and
Gadun et al. 2000).
To model solar convection realistically, requires  
a realistic equation of state, realistic opacities and proper  treatment of radiative transfer in the 
shallow layers. SN and KC are the two most freqently cited 
3D models of this type of stratified convection. As their  approaches differ considerably, 
both in numerical methods and 
in input model physics, it is important to find out, what particular 
aspect of the respective simulations, caused their results to be different.

The aim of this paper is to describe 
solar surface convection which not only has the KC realistic physics, but also  
has a realistic (as is presently possible) geometry.
To achieve  this we had to increase both the depth and width of the original  KC model, 
until the side walls and the horizontal boundaries had only a minimal effect on the flow.
The simulations themselves model a region less than a few thousand
kilometers in depth, as measured inwards from the visible solar surface. 
In the deep regions of the solar convection zone, the turbulent 
velocity is subsonic and the superadiabaticity is close to zero.
However, 
within a few hundred   kilometers  of the solar surface, the 
convective  flux starts to decrease. 
 As the total flux is constant,
the radiative 
flux must increase to offset the drop in the convective flux.
This is achieved by a
 rise in the local temperature gradient $\na$.
This region of inefficient convection is 
called the superadiabatic layer (SAL).  In the SAL, the superadiabaticity $\na -\na_{\rm ad}$
is positive and of order unity (Demarque, Guenther \& Kim 1997, 1999).
As the  buoyancy force is large in the SAL,
the region is characterised by highly turbulent velocities 
and large relative thermodynamic fluctuations.
The turbulent velocity also gives rise to a significant turbulent pressure
(approximately 15 \% of the gas pressure). This  moves out the convection surface,
modifying the SAL and the stratification. 
In one-dimensional (1D) models of the solar convection zone based on the mixing
length theory (MLT)(B\"ohm-Vitense 1958), the  velocity is set to zero above the
convection boundary. 
However, three-dimensional (3D) numerical simulations
described in Cattaneo et al. (1990), 
have shown that just above the top of the convection layer 
the turbulent velocities are still 
high.

There are several motivations for
such  simulations  among stellar physicists.  One
is to understand the effects of turbulence
on the structure of the outer solar layers as revealed by the observed
frequencies of solar {\it p}-modes.  Another is to explain the excitation
mechanism of the {\it p}-modes. Still another is to derive more realistic
surface boundaries for stars with convection zones from first
physical principles, free of the arbitrary assumptions of the
mixing length theory. And finally, such simulations may be of help in  
investigations of the solar dynamo.

Over the last few years, the science of helioseismology 
has  provided some very precise  measurements of the
{\it p}-mode oscillation frequencies   (to within one part in a thousand) in the surface layers 
of the Sun \cite{harvey1996}. The discrepancy between 
the observed {\it p}-mode frequencies and
those calculated from solar models, is known to be primarily due to the approximations
made in modelling the surface layers, where
turbulent
and radiative losses are significant (Balmforth 1992; Guenther1994).
In the 3D simulation described in  KC, the turbulent pressure pushed the convective boundary 
radially outwards from its original position (that was computed using the MLT). 
This situation  was mimicked in the 1D solar model 
by tweaking  the opacity in the
outer layers  \cite{demarque1999}. 
This resulted in improved {\it p}-mode frequencies for low and 
intermediate degrees.
However, Demarque et al. also showed that the mixing length prescription of 
Canuto \& Mazzitelli (1991), which had 
a completely different SAL structure than   KC, 
could produce a similar improvement of the {\it p}-mode frequencies for the 
same $l$-values.
Full details of the different approaches,
the contrasting  SAL structures and the resulting {\it p}-modes, are described  
in Demarque et al. (1997, 1999).
Later, Rosenthal et al (1999)  used another approach to compute the {\it p}-mode 
frequencies. 
These authors 
patched the mean stratification (horizontal average) of a 
3D hydrodynamical simulation,
onto a 1D MLT envelope model.
In order to get a smooth 1D model, they adjusted the mixing length and the 
amplitude of the turbulent pressure to match the 3D simulation. 
The computed frequencies were closer to  
the observed frequencies than if a standard solar model is used, thus 
showing the importance of including turbulence in modeling the outer layers of
the Sun.
More recently,
Li et al. (2002) found similar results to Rosenthal et al. by 
inserting  
the averaged turbulent 
pressure and turbulent kinetic energy  
directly into the 1D stellar model. 
Their method is  
applied to two of the simulations described in this paper (see section 4.3).
As the turbulent kinetic energy and turbulent pressure were 
very small at the base of the 3D model, they were  set to zero 
in regions of the 1D stellar model that lay below the 3D model domain.  
This 
required the usual adjustment of the mixing length 
parameter and the helium abundance to calibrate the perturbed stellar model.
No other adjustable parameters were employed. 
The improvement in the eigenfrequencies was found to be  
primarily  due to the inclusion of
turbulent kinetic energy flux in the 1D stellar model.

After extensive testing,
we found that our simulations are in good agreement with other
numerical studies  of the surface layers (e.g. Rosenthal et al. 1999, Asplund et al. 2000).
In addition, by
incorporating the computed 3D turbulence into a  1D stellar model,
we were able  to produce
solar surface eigenfrequencies ({\it p}-modes) that were very close
to the observed frequencies.
As our eventual goal is to simulate the
SAL in stars other than the Sun, 
it is vital to be sure we are modelling the Sun as correctly as possible.

\section{Modelling realistic solar surface convection}
\label{3d}
To model surface layer convection  in the Sun
as realistically as possible,
we take the following approach:
\begin{enumerate}
\item
Using  a stellar evolution code (YREC e.g. see Guenther et al. (1992)),
we compute a standard stellar model from which the initial
density $\rho$ and  internal energy $e$, required by the 3D simulations, are derived. 
From  an arbitrary  initial velocity field ${\bf v}$ ,
we  then compute
$\rho$, $E (=1/2 \rho v^2 + e)$, $\rho v_x$, $\rho v_y$ and $\rho v_z$.
These are the dependent variables of the governing equations.
The horizontal directions are $x$ and $y$, and $z$ is radially outwards.
\item
 Using the same tables for the equation  of state and the opacities as in the stellar model,
we then compute the pressure 
$P(\rho, e)$, temperature $T(\rho, e)$, Rosseland  mean opacity  $\kappa(\rho,e)$, specific heat capacity
at constant pressure  
$c_{\rm p}(\rho,e)$, adiabatic gradient 
$\na_{\rm ad}(\rho,e)$   and some thermodynamic derivatives.

\item
The radiation flux is then computed using the diffusion approximation in the
optically thick regions  and the 3D Eddington  approximation in the optically
thin layers.
\item
We then integrate the  Navier-Stokes equations over  one time step to compute
a new set of dependent variables and return to (ii).
\end{enumerate}                   
\subsection{Realistic initial conditions: steps (i) and (ii)}
The solar model  uses the same realistic physics as described in
Guenther \& Demarque (1997).
In particular, the low temperature opacities of Alexander \& Ferguson (1994)
and the OPAL opacities and equation of state were used \cite{iglesiasrogers1996}.
Hydrogen and helium ionisation, and the diffusion of
both helium and heavy elements are included.

In the initial model, the atmospheric layers, which are in radiative equilibrium,
are assumed to be gray.  Deeper in, in the convectively unstable region,
the thermal structure is described by the MLT, which prescribes the
temperature gradient $\na$. 
In the original KC simulation, the Eddington approximation T($\tau$) relation was 
used in the atmosphere. 
In this case, the values of the parameters, X,  Z and $\alpha$, in the 
calibrated Standard Solar Model (SSM)  
are 
$(X,Z,\alpha) = (0.7385,0.0181,2.02)$,
where X and Y are the hydrogen and helium
abundances by mass, and $\alpha$ is the ratio of mixing length to pressure scale
height in the convection zone, required to match precisely the solar radius.  It should
be mentioned that although the values of X and Y in the calibrated SSM depend little on
the treatment of the surface layers, the value of $\alpha$ is sensitive to the choice
of T($\tau$) relation in the atmosphere (Guenther et al. 1992).

We note that for simulations discussed in this paper (those listed in the appendix), 
the  empirical  
Krishna-Swamy  $T(\tau)$ relation for the Sun \cite{krishnaswamy1966} was used
instead
of the Eddington approximation.  
The same model as KC, but constructed with the Krishna-Swamy  T($\tau$) relation, yields  
 $(X,Z,\alpha) = (0.7424,0.01706,2.1319)$ (this is model KC2 in the appendix).
Because of the inclusion of helium and heavy element diffusion during the evolution,
the initial X and Z (denoted by X0,Z0) were slightly different. 
For KC they were (X0,Z0)=(0.7066, 0.0201), 
and for KC2 they were (X0,Z0)=(0.710,0.019).
The difference in initial atmospheric structures between simulations KC and KC2 provides 
us with the opportunity to verify that   
that the final state in the simulations is not strictly dependent on the 
choice of T($\tau$) relation in the initial model.  This important test is discussed 
in the appendix. 

For each timestep of the numerical integration we need to solve the
complex equation of state (step (ii)). This adds considerable computational time compared to
an ideal simulation,  such as with a perfect
gas.
The 3D hydrodynamical simulations use identical opacities and equation
of state as used in the 1D reference SSM, which served as the initial model.
The side boundaries are  periodic, while
the
top and bottom  boundaries are stress free.
A constant heat flux flows through the base, and the top is a perfect conductor.
To ensure that mass,  momentum and energy are fully conserved,
we use impenetrable (closed)
top and bottom boundaries.
 
A particular model is specified by $g$ (the surface gravity)
and $T_{\rm eff}$ (the effective temperature). 
Aside from the viscosity
coefficients there are no other free parameters. 
\subsection{Radiative  transfer: step (iii)}
In the SAL,
the photon mean free path  may not be small enough to use the diffusion
approximation. Consequently one is forced to either solve the full radiative transport equation or
consider the three-dimensional Eddington approximation (Unno \& Spiegel 1966), which is a higher order approximation
than the diffusion approximation, and is valid in the optically thin regions.
Computationally, to solve the full radiative transport equations is formidable, 
and only a few ray directions are
currently used in this approach (SN). In our simulations, 
we have chosen to use the three-dimensional  Eddington
approximation. 
 
In the deeper part of the domain ($\tau > 10^4$), we use
the diffusion approximation,
\be
Q_{\rm rad} = \na \cdot \left[\frac{4acT^3}{3\kappa\rho}\na T \right],
\ee
where $\kappa$ is the Rosseland mean opacity,
$a$ is the Boltzmann constant and $c$ is the speed of light.
In the shallow region
$Q_{\rm rad}$ is computed as
\be
Q_{\rm rad} = 4 \kappa \rho (J - B)
\ee
where the mean intensity $J$ was computed by using the generalized three-dimensional Eddington
approximation (Unno \& Spiegel),
\be
\na \cdot \left( \frac{1}{3\kappa\rho} \na J \right) - \kappa  \rho J + \kappa \rho B = 0,
\ee
where $B$ is the Planck function.
This formulation is exact for isotropic radiation in a gray atmosphere,
and without requiring local thermodynamic equilibrium,
the Eddington approximation describes the optically
thick and thin
regions exactly (Rutten 1995). 
To study spectral line profiles and spectral energy distribution, requires frequency
dependent radiative transfer.  However, in a study  of the SAL, to
serve as a surface boundary condition for stellar models or for
comparison with results of helioseismology, we found a gray atmosphere
to be adequate.
\subsection{Hydrodynamics: step (iv)} 
\label{les}
For deep ($v << c_{\rm s} $, where $v$ is the flow velocity
and $c_{\rm s}$ is the isothermal sound speed)  and efficient  ($\na -\na_{\rm ad}$ just above zero) convection, Chan \& Sofia (1989) showed that
the MLT is a very good approximation to the
real situation.
However, in the SAL, both  $v/c_{\rm s}$
and $\na -\na_{\rm ad}$,
can be of order unity. In this case, the MLT is unlikely to apply. The validity of this argument
is confirmed by helioseismology.
The run of the sound speed in the SAL derived by inversion of the helioseismic data does in fact
disagree with the MLT model (Basu \& Antia 1997).  
In such an environment, the governing  hydrodynamic equations are the fully compressible
Navier-Stokes equations (see for example  Kim et al. 1995).
\bea
\pa \rho / \pa t &=& - \na \cdot  {\bf \rho v }\\
\label{ns1}
{\bf  \pa \rho v} / \pa t & =& - {\bf\na \cdot \rho  v v}
- \na  P
+ {\bf \na \cdot \mbox{\boldmath$ \Sigma$ }}
+ \rho {\bf g } \\
\label{ns2}
\pa E /  \pa t& =&   - \na \cdot [(E+P) {\bf v
-  v \cdot \mbox{\boldmath$ \Sigma$ }}
+ {\bf f} ]\\
&& + {\bf \rho v \cdot g} + Q_{\rm rad}\nonumber
\label{ns3}
\eea
where $ E = e + \rho {v^2} /2$ is the total energy density and $\rho, {\bf v }, P, e$
and ${\bf g}$, are the density,  velocity, pressure, specific internal energy
and acceleration due to gravity, respectively. $Q_{\rm rad}$ is the energy 
transferred by radiation (see previous section) and ${\bf f}$ is the diffusive flux.
Ignoring the coefficient of bulk viscosity, the viscous stress tensor
for a Newtonian fluid
is $\Sigma_{ij}=\mu(\pa v_i/\pa x_j+\pa v_j/\pa x_i)-2\mu/3(\na \cdot {\bf v})\delta_{ij}$,
where $\mu$ is the dynamic viscosity and $\delta_{ij}$ is the delta function. 

In fully developed turbulence the ratio of the length of the largest eddy
to the dissipation length  is  ${\rm Re}^{3/4}$, where Re is the
Reynolds number, (Landau \& Lifshitz 1987).
In the Sun, Re is the order of $10^{12}$ which means
about $10^9$ scales per dimension. A 3D direct numerical simulation
of the Sun would thus require about $10^{27}$ grid points !
 
The Large Eddy Simulation (LES)  approach assumes that the small scales
are independent of the resolved scales (large eddies)  and can be
parameterised as a diffusion process.  In this case $\mu$ is 
an eddy viscosity defined in terms 
of the resolved velocity  (Smagorinsky 1963),
\be
\label{sgs}
\mu=\rho(c_\mu\Delta)^2(2 \mbox{\boldmath $\sigma : \sigma$})^{1/2}.
\ee
The colon inside the brackets denotes tensor contraction of the rate of strain tensor
$\sigma_{ij} = (\na_i v_j+\na_j v_i)/2$.
The  sub-grid scale  (SGS) eddy coefficient $c_\mu$, is set to 0.2, the value for incompressible
turbulence, and $\Delta = (\Delta_x \Delta_y)^{1/2}\Delta_z$ is an estimate of the local mesh size.
To handle shocks, $\mu$ is
multiplied by $1 +C \cdot (\na \cdot {\bf v})^2$, where
the constant $C$  is made as small as possible, while still  maintaining numerical stability.
As $\mu$ is dependent on the velocity  divergence, any
large velocity gradients are smoothed out by the increased viscosity. If a shock
occurs it is not resolved, but smeared out by a local increase in viscosity.

The diffusive flux ${\bf f} = -(\mu / {\rm Pr}) T \na S $, where the horizontal mean of the 
entropy gradient $\na S \leq 0$
i.e. the convection zone,
and ${\bf f} = -(\mu c_{\rm p}/{\rm Pr}) \na T $ 
where  the horizontal mean of $\na S \geq  0$ 
i.e. the radiation zone (the Prandtl number Pr  is defined below).
In the convection layer the SGS diffusive flux tends to smooth out entropy fluctuations
and make the layer close to adiabatic (in analogy with turbulent mixing).
The change in the form of the  diffusive flux above the convection boundary is
necessary  because  the  SGS's should  continue to transport heat radially outwards though the top of the box.
Away from the horizontal boundaries, ${\bf f}$ is close to zero.
At the base, ${\bf f}$ is equal to $\sigma {T_{\rm eff}^4}$, 
where $\sigma$  is the Stefan-Boltzmann  constant.  
The Prandtl number Pr $=\nu / \kappa $, where $\nu$ is the kinematic viscosity and
$\kappa$ is the thermal diffusivity. In the simulations Pr =  1/3.
Due the inclusion of radiative energy transport
the effective Pr is actually much smaller
and not constant.        
                                                               
\section{Numerical integration: obtaining accurate statistics}
To simulate the highly stratified SAL of the Sun we need
to relax the initial layer and then compute accurate statistics.
The former requires  a long computation, while the latter a small timestep.
In compressible hydrodynamics however, with an  explicit numerical method,
the timestep must be less than the time for a sound wave to traverse two adjacent grid points.
This is known as the Courant-Friedrichs-Levy (CFL) stability criterion.
Because of these considerations  
the simulations 
were done in two stages.

Firstly, using an implicit code in which the
time step is restricted by the flow speed rather than the  sound speed, the initial 
hydrostatic layer was allowed to adjust its thermodynamical structure until 
it was close to hydrodynamic equilibrium. This {\it thermal adjustment} phase took at least 5 hours of 
solar surface convection time. 

Secondly, using a second order accurate explicit code, quantities were  averaged over
a time that was long enough for the averages to be independent of the 
integration time. As the explicit timestep is about five times smaller 
than the implicit timestep, prior to statistical averaging, the code was run 
for a few thousand  timesteps. This  allowed the simulation to adjust to the new timestep.
The {\it statistical convergence} 
took  at least  an hour of solar surface convection time.

\subsection{Thermal adjustment}

The implicit code was the Alternating
Direction Implicit Method on a Staggered grid (ADISM) developed by Chan \& Wolff (1982).
This code was used to relax the fluid to a self consistent
thermal equilibrium.  

The entire layer was assumed to be relaxed when 
\begin{itemize}
\item 
the energy flux leaving the
top of the box was within 5 $\%$ of the input flux at the base;
\item
the horizontally averaged vertical mass flux was less 
than $10^{-4} $ g/${\rm cm}^2/{\rm s}$  at every vertical level; 
\item
the overall thermal structure 
did not 
change much over time;
\item
and the maximum velocity in the box was roughly constant. 
\end{itemize}
These  criteria must be satisfied, before 
any useful statistical data can be gathered.

\subsection{Statistical convergence}
A second order explicit method (Adams-Bashforth time
integration) gathered the statistics of the time averaged state. This code is much more accurate
than ADISM, for example the  mean energy flux leaving the
top of the box was within 1 \% of the input flux at the base. 
On a 667 MHz Alpha processor, each integration step on a  $80 \times 80 \times80 $ grid, 
required about 5 seconds of CPU time.   

The time required for statistical convergence  depends on the
particular quantity being averaged. 
Conserved quantities converge very fast.
For example,  the horizontally averaged vertical mass flux was less than
$10^{-5} $ g/${\rm cm}^2/{\rm s}$ after 5 minutes of solar time integration. 
On the other hand,  rms velocities
converged in  a few  eddy turnover times (at least  30 minutes of solar time).
While second order turbulent quantities, such as the horizontal Reynolds stress,
took  even  longer to converge (at least 80 minutes of solar time). 
\subsubsection{Some statistical  definitions}
In
a turbulent fluid a quantity $q$ can be split into a mean and a fluctuating part,
\be
q = \overline{q}(z)+ q'(x,y,z,t).
\ee
The overbar represents a  combined horizontal and temporal average, i.e.
\be
\overline{q}(z) = \frac{1}{t_2-t_1} \int\limits_{t_1}^{t_2} \left(
 \frac{1}{
 (L_x L_y)
}
\int q   dx dy   \right)dt.
\ee
$t_1$, is a time  after the system has reached a self-consistent
thermal equilibrium (the thermal adjustment time). $L_x$ and $L_y$ are the 
horizontal widths of the box in the x and y direction respectively. 
The time required for statistical convergence is $t_2-t_1$.

The rms value of a quantity $q$ is defined as 
\be
q''=\overline{q^2}-{\overline{q}}^2,
\ee
while the   correlation coefficient of two quantities $q_1$ and $q_2$,   
is defined as 
\be
 C[q_1'q_2']=\frac{\overline{q_1 q_2}-\overline{q_1}\hspace{1mm}\overline{q_2}}{q_1''q_2''}x .      
\ee

As the  simulations have  periodic side boundaries,
symmetry requires that
\begin{itemize}
\item
$C[v_x' v_y'] = 0 $ 
\item
$v_x''=v_y''$
\end{itemize}

\begin{figure}
\epsfig{file=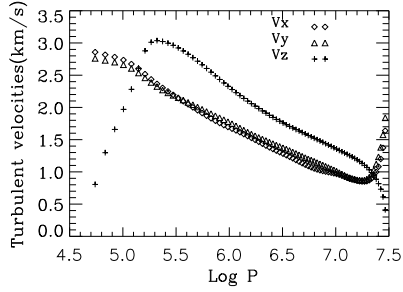,width=6.cm}
\caption{\label{f2}The rms turbulent velocities in the horizontal 
and vertical directions versus depth. The closeness of the two horizontal velocities confirms
that the simulation is close to statistical convergence.}
\end{figure}

The run of $v_x''$ and $v_y''$ after 
80 minutes of time integration is shown in Fig. \ref{f2}.
The closeness of the horizontal velocities confirms that  the simulation 
is close to statistical convergence. 

\begin{figure}
\epsfig{file=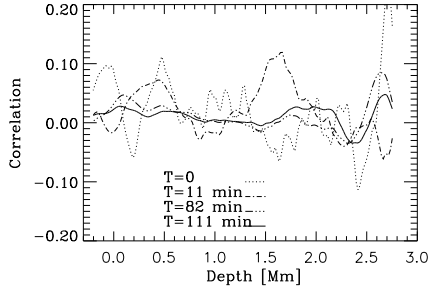,width=6.cm}
\caption{\label{f3}$C[v_x' v_y']$ for three integration times.
As $C[v_x' v_y']$ converges more slowly than most other quantities, the degree
of statistical convergence can be estimated by how close $C[v_x' v_y']$  is to zero.
Convergence is slowest near the bottom of the box.}
\end{figure}

By examining many different
simulations we found that 
the run of $C[v_x' v_y']$ is generally a much  stricter test of convergence.
Fig. \ref{f3} shows $C[v_x' v_y']$ measured for  4 different integration times.
Even after 80 minutes, $C[v_x' v_y']$ is still about 0.1 near the bottom.
Convergence is faster in the upper layers because when
the convecting fluid elements  move up through the stratification, their rapid expansion
smoothes out small scale fluctuations.

\section{Main results}

\subsection{Effect of domain size}
\label{numericaltests}

After substantial numerical testing  which is described in detail in the appendix, 
we found that :
\begin{enumerate}
\item
An undesirable  effect of the impenetrable horizontal boundary at the bottom was too
speed up the
overall flow. This produced an artificially high convective flux.
Just below the surface the radiative flux is a significant fraction of the total flux.
As the total energy flux is fixed, to accommodate the increased convective flux, 
the radiative flux had to reduce. 
This 
was achieved by an (unphysical)  drop in the temperature gradient $\na$ in the SAL region. 

To avoid
this, the lower  boundary has to be positioned far enough away from  the
surface, so that the velocity at the base is small and uncorrelated from the
motions near the surface.
\item
If the width of the box was too small then the turbulent kinetic energy of
the granules was artificially small. This is because the movement of the larger granules
was restricted by the walls of the box. 

To avoid this, the aspect ratio was doubled  until
the turbulent kinetic energy was unaffected by a further increase in aspect ratio.
For the Sun this required a width of 2.7 Mm.
\end{enumerate}                           
\subsection{Comparison with previous 3D numerical simulations}
\subsubsection{Vertical velocity}

Using the SN code, Asplund et al. (2000) computed a series of simulations 
of solar surface convection in   a domain with a  depth   of 
about 4 Mm and width  of 6 Mm. 
The best resolution in Asplund et al. was $200^2 \times 82$.
The rms vertical velocity and the mean velocity 
are shown in 
Fig.s 3 and 4 in their paper. Corresponding velocity 
plots from our best model (model C in the appendix) are shown in Fig. \ref{vcomparison}.
Note that we have radially upwards as
the positive direction,
the opposite to the Asplund paper.
Despite all the differences between the SN and KC approaches,
away from the boundaries, the rms and mean vertical velocities  in
our best model are very similar to those in the  
best Asplund et al. simulation. The most  noticeable differences occur
at the top
and bottom because in our simulations the
vertical velocity  is forced dropped to zero.
In Asplund et al.
the
transmitting boundaries allowed the velocity to decrease more gently.  

\subsubsection{Superadiabaticity}
The superadiabaticty  $\na - \na_{\rm ad}$   
for the MLT, and models KC2 and C,  
are plotted in Fig. \ref{salr}. The abscissa is the  radius of the simulation 
divided by the radius of the Sun.
For model C, the superadiabaticity has a maximum of about 0.6, which  is close to the
value given by the SN code (see Fig. 3 in Rosenthal et al.). 
The position of the top of the convection layer (as determined by the Schwartzchild criterion)
was pushed out further in KC2 than in C. This is  because of the higher turbulent pressure.

When  the SAL was moved outwards 
the convective efficiency was reduced and radiation was forced to carry more 
of the total flux. This resulted in an increase  in the height of 
the SAL in C (triple dot dashed line),  compared to the MLT (crosses). 
However, in KC2 (dashed line) the convection was speeded up by the 
lower boundary and thus is (incorrectly) more efficient than the MLT. 
This resulted in a  drop in the height of the peak of the SAL 
compared to the MLT.

\subsection{Comparison with observational results: {\it p}-mode oscillation frequencies}

\subsubsection{Implementation
of 3D turbulence into 1D stellar models}

By analogy with the work of Lydon \& Sofia (1995) on magnetic effects,
the 3D turbulence  is parameterised in terms of two quantities (Li et al. 2002),
the turbulent kinetic energy per unit mass,
\be
\chi = \frac {1}{2} {v''}^2 
\ee
where ${v''}^2 = { v_x''}^2 + {v_y''}^2+{v_z''}^2$ is a dimensional quantity,
and an anisotropy parameter,
\be
\gamma = 1+2(v_z''/v'')^2.
\ee
The $z$ direction
is parallel to the radial direction.

The turbulence is included  by incorporating   
2 new variables $\gamma$ and $\chi$ into the 1D stellar model.
To understand how this is works, consider a perfect gas in which 
the ratio of the specific heats $\gamma=c_{\rm p}/c_{\rm v}$, the  internal energy $e= c_{\rm v} T$ and
the gas pressure $P=\rho R T$.  The quantities  $c_{\rm v}$ and $R$ are for unit mass of a gas and 
$T$ is the temperature.
The previous three equations can be expressed as, 
\be
\gamma=1+P/(\rho e)
\ee

If we replaced gas quantities by turbulent quantities, i.e. 
$P$ by $P_{\rm turb}(=\rho {{v_z}''}^2)$ and $e$ by
$\chi$ and rearrange, 
then we would get:
\be
P_{\rm turb}=(\gamma -1) \rho \chi.
\label{pturb}
\ee
where $\gamma$ is defined in terms of 
turbulent quantities (i.e. rms velocities). 
Including $\chi$ and $\gamma$ is  
{\it equivalent} to including $\chi$ and $P_{\rm turb}$ in the  
1D stellar model. The two variables $\gamma$ and $\chi$
are included in the mathematical system in a self consistent manner (see 
Li et al. for the full mathematical treatment)
 
For example, the equation of state for the 1D model becomes,
\bea
\rho=\rho(P_T,T,\chi,\gamma) \nonumber
\eea                                                                       
where $P_T= P_{\rm gas}+P_{\rm rad}+P_{\rm turb}$.
\\
The continuity equation and the equation of transport of energy by radiation
remain the same regardless of turbulence.
In terms of $\gamma$ and
$\chi$, the hydrostatic
equilibrium is,
\be
  \frac{\pa P_T}{\pa M_r} = - \frac{GM_r}{4\pi r^4} - \frac{2(\gamma-1)\chi}{4\pi r^3}.
\ee
where  $M_r$, $G$ and $r$ have their usual standard stellar model (SSM) definitions. 
The energy conservation equation,
\be
\frac{\pa  L_r}{\pa M_r}=\epsilon - T \frac{dS}{dt},
\ee
is also
affected by the inclusion of $\chi$,
as 
\be
TdS = dU +(P_T-P_{\rm turb})d(1/\rho) + d \chi,
\label{2ndlaw}
\ee
where $d \chi$ represents the work done by turbulence.  
The quantities $L_r$ and $\epsilon$  have their usual SSM definitions.

The convective flux in the 1D model $F_{\rm conv} = \rho T DS \hspace {1mm} V_{\rm conv}$,
now includes the turbulent kinetic energy flux $\chi V_{\rm conv}$ (without turbulence
$\chi=0$, so $d \chi = \chi$). $V_{\rm conv}$ is the MLT convective velocity and DS is the entropy excess.
Equation \ref{2ndlaw} is  the `non-standard form' of the 2nd law as described in
Lydon \& Sofia.
As turbulence is in  general anisotropic it is wrong to treat
the work done by turbulence as $-P_{\rm turb}dV$. This is essentially  because
$P_{\rm turb}$ is a tensor
and $P_{\rm gas}$ is a scalar. 
                                          
The equation of energy transport by convection,   does not change in form, but $\na$
is different from that without turbulence. The
equations that govern envelope integrations are   changed
accordingly (see  Li et al.).

\subsubsection{Solar {\it p}-mode oscillation frequencies}
 
To investigate the effect of turbulence on the solar {\it p}-mode oscillation
frequencies, the runs of $\chi$ and $\gamma$ for   models KC2 or C  
were incorporated into the 1D stellar model using the Li et al method.
For each run, the {\it p}-mode frequencies for $l=0,1,2,3,4,10...100$, were computed using Guenther's
pulsation code, (1994), under the adiabatic approximation.
When comparing the computed and observed eigenfrequencies, one should really
include the proper modeling of radiative gains and losses by
computing the non-adiabatic frequencies (Guenther 1994).
However, to help isolate the effects of turbulence on the
{\it p}-mode frequencies, we  computed  the adiabatic frequencies.
The difference between observed and computed adiabatic {\it p}-mode frequencies
for the standard solar model  is shown  in Fig.  \ref{pmodessm}.  The frequency difference is
scaled by the mode
mass $Q_{nl}$ (e.g., Christensen-Dalsgaard \& Berthomieu 1991).

The $Q_{nl}$ weighting attempts to correct the {\it p}-mode frequency
differences, by removing the dependency on mode inertia. The  mode
differences with high mode inertia  get more weight than those with lower inertia.
This tends to give greater weight to the low $l$, deep penetrating
modes and less significance to the very high $l$ shallow modes.  
The result is the removal of  the $l$ dependence in 
the frequency differences due to perturbations 
in the near  surface regions. We are left with only $n$ dependence,
which is equivalent to frequency dependence.  The $Q_{nl}$ weighting 
enables us to  
see that the discrepancy  between
the observed
and the computed adiabatic frequencies is worse at the high frequencies,
thus pointing to a problem in the surface layers.
Without the  $Q_{nl}$ weighting, the mismatch would appear to varying
degree, at all frequencies and therefore it would not be as easy to claim
that the outer layers are responsible for the mismatch.

In the statistically steady state (after the flow is thermally relaxed),
the turbulent kinetic energy   flux
should be  proportional to the  energy input rate at the base.
The turbulent kinetic energy $\chi$ should be independent of the geometry of the computational domain.
If the box is too shallow  or the width too narrow,
then $\chi$ can be wrong.
To illustrate the effect of the geometry on the frequencies, $\gamma$ and $\chi$ from the KC2 model was used 
to compute a stellar model. The computed adiabatic
frequencies with turbulence  derived from KC2 are shown in Fig. \ref{pmodeshallow}.
The resulting frequencies are much worse than  the standard solar model.
However, if we use model C instead of KC2, then
the derived frequencies were  improved
considerably (Fig. \ref{pmodedeep}).
The difference between the computed and the observed frequencies is  less than 5 $\mu$ Hz.
The inclusion of turbulent kinetic energy in 1D models
in addition to turbulent pressure, has a similar effect on the adiabatic frequencies
to the inclusion of radiative losses/gains (computation of non-adiabatic
frequencies). Both reduce the difference between the adiabatic frequencies  and the
observed frequencies
by an order of magnitude. This mean that the turbulent correction to the {\it p}-modes
is as important as radiative losses/gains.
                         
\subsection{Characteristics of the solar  granules} 
\subsubsection{Size}

Fig. \ref{granr} shows a  set of contours of the instantaneous vertical velocity. From top to bottom
the frames depict  depths of 
-0.14 Mm, 0.03 Mm, 0.2 Mm and 1.0 Mm.
The depth was measured positively 
inward from the visible solar surface.
The contours themselves have  been  derived from  a simulation of $80^3$ grid points, in a domain 
with  a horizontal area of 3.75 Mm $\times$ 3.75 Mm, and a depth of 2.5 Mm. 
At this instant in time there appear to be between 3 and 4 granules near the surface. 
The thick black lines  represent  the strong downflows which occur at the sides of the granules. The lighter 
regions denote upflowing fluid or weak downflows.
The  granular 
pattern seems to persist over the first three frames, suggesting  that the granule structure is 
correlated over most of the SAL (between 0 Mm and  0.25 Mm). By the fourth 
frame  there is not much sign of granulation. 
As the last  contour is still more than 2 pressure scale heights above the base,
the  break up of the granules  is probably not  due to the impenetrable bottom boundary.
In other words the bottom boundary is far enough away from the surface.
Furthermore, as the shape of the granules does not appear to be influenced by the side walls,
the width of the box is also big enough.

The thin closely spaced dark vertical/slanted parallel
lines that appear in the first contour, are signs of two grid waves.
This is an indication of insufficient damping and  
is only seen at the very top of the box.
It is due to the low SGS viscosity, which is proportional to 
density ((equation \ref{sgs}).
This is  is an  undesirable   numerical effect. But 
as the oscillations  occur only at the very top of the
box, where the density is very small, their effect on the
convection below is minor.
Before starting  the time integration,  
we  were able to damp out such oscillations, by slightly 
increasing the viscosity near the top.
As these contours only  represent one instant in time,
such a picture can only provide a limited
idea of the nature of the solar  granulation. If the same contours are computed 
at a later time, a completely different picture  would be  seen. 
For example the third frame of figure 22  is the same simulation, but the contour was computed 2 solar minutes later. 

Clearly any useful data can only be derived  from long time averaged statistics.
One  characteristic vertical length scale associated with convective turbulence is the 
half-width of the 2-point vertical velocity correlation C[$v_z' v_z']$. If we assume convective eddies 
to have an aspect  ratio of unity, then this length scale 
gives us an idea of the size of the eddies in a turbulent fluid.
Using  simulation D (described in the appendix), we computed C[$v_z' v_z']$ at a series of depths 
from the top of the box inwards.   
The results are shown in Fig. \ref{hw}.
From about  0 Mm  to  1 Mm the eddy size remains  
close to 600 km. From the begining to the end of the plateau, both density and pressure 
have increased by an order of magnitude. 
Over this range the eddy does not seem to be affected by stratification.
However, over the next  Mm  the eddy size increases to about 1100 km. After that 
the lower boundary starts to influence the eddies. The bends  
at the far left and far right of the plot  signal the approach of the upper and lower 
boundaries.  

The velocity contours and  the velocity correlation length both suggest that the granules have some 
coherent vertical structure. The contours show intergranule lanes that do not shift much 
between  0 and 0.2 Mm, i.e. the granules are like cylinders. The half-width suggests a vertical 
size of about 600 km, between 1 Mm and the surface. This  length scale does not seem to be 
affected by the stratification. The strong vertical coherence implies that the 
convection is  dominated by strong downflows that originate close to  the surface. 
SN showed that solar granulation is primarily driven by radiative cooling at the surface.
When ascending fluid approaches the surface, it looses heat so rapidly that very strong 
downflows are created. The downflows are not deflected by the surrounding fluid
until they are about 1 Mm below the surface, by which point they have weakened 
enough  to be deflected/broken up 
by the surrounding fluid motion. 
\subsubsection{Heat transport}

In Chan \& Sofia (1987,1989) the 2-point vertical velocity and temperature  
correlations were found to scale with 
pressure scale height rather than density scale height. 
This study was for deep efficient convection. 
In shallower layers Kim et al. (1995) found that 
while the vertical velocity correlation scaled with 
both density and pressure scale height,  the 
temperature could not be scaled with either.
The difference from  Chan \& Sofia's results, 
was caused  by the inclusion of 
the coupling of the partial ionisation  
with convection (they were treated separately by Chan \& Sofia). 
As a fluid parcel moves upwards through regions of 
decreasing  ionisation, it  liberates ionisation energy which 
increases the  buoyancy of the fluid parcel. 
However, because radiation was modelled  by the diffusion approximation (which
is known to break down at some point in the  SAL),
the Kim et al. simulations could only include the lower half of the SAL.

The present models include all of the SAL.
Figs.  \ref{vzvz} and \ref{ss} show $C[v_z' v_z']$  and  $C[S' S']$
at  depths of 0.06 Mm, 0.2 Mm, 0.3 Mm, 0.47 Mm and 1 Mm, denoted by
solid, small-dashed, dot-dashed, triple-dot dashed and long-dashed
lines, respectively. The curves have been centered about
their respective maxima. As the plots do not coincide, neither 
quantity scales with the pressure scale height.
Furthermore, if density is used rather than pressure, there was no
noticeable improvement (not shown).   

The width of the $C[v_z' v_z']$ distribution  
decreases between 0.06 Mm  and  1 Mm. This is because the width of the geometric 
distribution  is roughly constant 
(i.e. the half-width is always about 600 km, as shown in the previous figure), but the scale height
increases inwards. 
The entropy correlation  looks very different. 
For $\Delta \ln P < 0$  the entropy 
correlation for the three most shallow depths drops 
rapidly, while for $\Delta \ln P > 0$ it shows a smoother 
decay.  This implies that ascending parcels near the surface,
lose heat very quickly, while descending 
parcels maintain their heat content appreciably longer.  
Over the last Mm before the surface, the eddies may have 
about the same  diameter, 
but they transport less and less heat 
as the surface is approached. 

This behaviour can be understood by 
examining the correlation $C[v_z' S']$ for  upflows  and downflows.
Fig. \ref{cvzs} shows that the granules are more efficient at transporting entropy downwards 
than  upwards. At the peak of the SAL, $C[v_z' S']$ 
is 0.9 for downflows  and 0.6 for upflows. 
This explains the one-sidedness of $C[S' S']$. 
Consider a fluid parcel 
that is initially just below the surface (say at a depth of 0.1 Mm). If that   parcel moves  downwards it will maintain 
its original entropy for about 1 PSH, whereas if the parcel moves 
upwards it it will loose its original entropy in 0.5 PSH. 
At greater depths $C[v_z' S']$ is similar in the upflows and downflows, hence 
$C[S' S']$ is more symmetric deeper down.
\subsubsection{A variable mixing length ?}

Over the last 1000 km before the surface, the eddies seem  to have
a nearly constant diameter of about 600 km. This 
would seem to imply a mixing length ratio (the correlation length divided by the 
pressure scale height) which increases towards the surface. 
However, as the surface is approached, radiation plays a greater and 
greater role in heat transport, so  
the mixing length ratio should  approach zero at the convection surface. 
This paradox can be partly resolved by noticing that as  the surface is approached, 
the eddies become less and less efficient at transporting 
heat upwards. This is shown in the plot of $C[v_z' S']$ for the upflows. 
A more reasonable candidate for the mixing length might  be 
the product of 
$C[v_z' S']$ and the velocity correlation half width.

\section{Summary and implications for solar models}

As we intend  to use our code to model the SAL in other stars, 
this paper is an important  benchmark for future  studies.
From these simulations of the Sun, we found that:  
\begin{enumerate}
\item 
The impenetrable lower boundary needs to be far enough away from the SAL 
so that by the time the fluid reaches the boundary, the velocities have become  
both weak and uncorrelated from 
velocities in the SAL. If the lower boundary is too close to the SAL, the kinetic energy will be 
overestimated.
\item
The horizontal cross section needs to  be large enough so that the 
side-walls do not restrict the movement of the granules.  
If the box width is too small then 
the kinetic energy will be underestimated.  
\item
There is a region close to the surface in which the vertical correlation length 
remains constant, even though the density and pressure vary by an order of magnitude.
\item 
In that region the mixing length theory, which assumes the correlation length to be a constant multiple of the 
pressure scale height, will not work.
\item 
The final equilibrium state is not strictly dependent on the initial model 
atmosphere (see appendix, section \ref{initial}).
\end{enumerate}

While these results  have  been found in a simulation of the Sun, 
it is reasonable to assume that similar criteria would apply to  the convection-radiation  
transition layers in other stars.  The effect of the boundaries should certainly 
be considered when simulating other convection-radiation  layers. The correlation length (half-width)  
seems to be  a very robust 
feature of the solar granules and could easily be measured in other computations. 
For example, in a recent proceedings (Robinson et al 2002)
we describe the application of this  model to the Sun at the sub-giant and the start of the 
red giant branch. Preliminary results indicate that, near the surface of each convection zone,  
the  ratio of the half-width to the stellar radius  depends directly on
the surface gravity.

\vspace{3mm}
This research is supported in part by NASA grant  NAG5-8406 (FJR, PD, SS).
YCK is supported by a Department of Astronomy, Yonsei University grant
(2001-1-0134). DBG's research is supported in part by a grant from NSERC of Canada. 
We also would like to acknowledge computer support from the  Centre
for High Performance Computing
at Saint Mary's University in Canada. 
Finally, we  thank H.-G. Ludwig  for helpful comments.
\section{Appendix : Numerical tests}
 
Numerical simulations of solar granulation need to  be robust.
Part of this means that the walls of the computational box should  not control the 
radiative hydrodynamics of the interior (e.g. the size of the granules, 
the structure of the SAL, the turbulent pressure, etc.). Nor should any
small changes in the 
initial conditions  affect the final equilibrium.
The following simulations were designed  to minimise these  uncertainties. 

The features of the individual simulations are summarised in Table \ref{models}.
In this table the geometric width in both the x and y directions 
is given in Mm, while the depth is given in 
Mm and as the number of pressure scale heights (PSH). Models A and B were 
created by vertically extending model KC2, while C, D and E were 
made by periodically extending  model B.
After periodic extension, the horizontal resolution in models D and E was halved.
Each time a new model was created, it was allowed to 
return to thermal equilibrium before the statistical integration  was begun.

\begin{table*}
\begin{minipage}{140mm}
\caption[Grid refinement study]{List of Simulations}
\begin{tabular}{ccccccc}
Model &Width (Mm)  & Depth (Mm) &Depth (PSH)&    $N_x \times N_y \times N_z$ & $\Delta_x$(km)& 
$\Delta_z$ (km)  \\
\\
KC/KC2 &1.35&0.9  & 4.4 & $60^3$      & 26 &17.5 \\
A & 1.35 & 2.1 & 7.2 & ${59^2}120$ & 26 &17.5 \\
B & 1.35 & 2.8 & 8.5 & ${59^2}170$ & 26 &17.5 \\
& &  & & & &  \\ 
C & 2.7 &  2.8 & 8.5 & ${116^2}170$ & 26  & 17.5 \\
D & 2.7 &  2.8 & 8.5 & ${58^2}170$  & 52  & 17.5 \\
E & 5.4 &  2.8 & 8.5 & ${114^2}170$ & 52  & 17.5 \\   
\end{tabular}
\label{models}
\end{minipage}
\end{table*}
For each model we computed three non-dimensional statistical quantities:

\begin{enumerate}
\item
The turbulent pressure
divided by the  mean gas pressure,
\be
{P_{turb}}^{*}= {\overline \rho} {v_z''}^2 / {\overline P},
\ee
where the `*' denotes a non-dimensional quantity.
\item
The
turbulent kinetic
energy per unit mass  divided by the 
isothermal sound speed squared,
\be
{\chi}^{*} = \frac {1}{2} \frac{{v''}^2}{{c_{\rm s}}^2}
\ee
where ${v''}^2 = { v_x''}^2 + {v_y''}^2+{v_z''}^2$ and $c_{\rm s} = \sqrt{{\overline P}/{\overline \rho}}$ 
is the isothermal sound speed.
\vspace{3mm}
\item 
The superadiabaticity  
\be
\na-\na_{\rm ad} = {\overline \frac{\pa \ln T}{\pa \ln P}}- \overline{\na_{\rm ad}}
\ee 
where $\na_{\rm ad}$ is computed  using  the OPAL equation of state 
\end{enumerate}
From now on we will drop the '*'
on $P_{\rm turb}$ and $\chi$, for convenience. 
All quantities in this section  are non-dimensional.

\subsection{The influence of varying the initial conditions}
\label{initial}
Here we compare the original KC simulation with simulation KC2 
listed in Table~1.
Both simulations have the same input physics (opacity tables, 
equation of state). 
As noted in section 2.1, the only difference between KC2 
and KC, is that the 1D initial models were based on 
slightly different model stellar atmospheres [i.e. the 
Eddington  T($\tau$) relation in KC, 
vs. the empirical  
Krishna-Swamy  $T(\tau)$ relation in KC2]. 
As the Krishna-Swamy relation  is from observations of the Sun, it was closer to the
final state of the solar simulation. 

Fig. \ref{f4} shows the initial hydrostatic (based on the MLT) and  
relaxed hydrodynamic runs of Log P vs Log T for the two simulations. 
The dotted and solid lines are the initial plots for 
KC and KC2, respectively. The horizontal and temporal averages of the relaxed 
states, are denoted by diamonds and crosses for KC and KC2, respectively. To demonstrate 
statistical convergence, the data of 
model KC2  was initially averaged over about 600 seconds 
of solar surface convection, whereas in  KC, the averaging time 
was 2240 seconds.  After 
a sufficient amount of time, KC2 (crosses) converged to  KC (diamonds).  
The MLT provides
the initial thermal structure, but the hydrodynamical turbulence
shapes the final structure.
Provided the initial conditions  are not too different, 
the hydrodynamics converges to the same equilibrium state. 

This is an important  preliminary  result. Even if the initial 
atmospheres 
are slightly different, the 
hydrodynamics  
dictates the final equilibrium. 
The thermodynamics of the resultant 3D simulation does not depend on 
the precise details of the MLT. The final equilibrium is not
determined by  the  exact value of the mixing length ratio in the stellar model, on
which the initial state of the simulation is based.  
The choice between the two  initial conditions in the atmosphere, 
does not affect the turbulent pressure,  
turbulent kinetic energy, or the superadiabatic temperature gradient. 
These quantities will depend on the geometry and grid resolution in the box.

\subsection{The influence of the lower impenetrable boundary}
\label{lowerbc}

The structure of the SAL in KC2 (or KC which is identical) 
is at odds with the SAL constructed with the Canuto-Mazzitelli (1991) approach 
(see Fig 1. in Demarque, Guenther \& Kim (1999)) or presented  in the  numerical 
simulation by Rosenthal et al. 
The most obvious difference is that KC2 predicts a broader SAL 
with a maximum of about 0.4, while Canuto-Mazzitelli predicts a peak of about 
unity and Rosenthal et al a peak of 0.6.
The height of the peak is related to the convective efficiency 
with respect to radiative heat transport.

Models KC2, A and B differ only in their domain depths.
Simulation A was constructed by adding  3 extra pressure scale heights 
to the base of model KC2 (see Table~1).  
We first thermally relaxed the lower layer, because
with only hydrostatic support (no turbulent pressure), the overlying layer
collapsed, and the simulation crashed. Also, when joining the
two layers, the turbulent viscosity was temporarily increased.
This smoothed out the fluctuations that were produced by suddenly removing
the lower boundary. Model B was  made by adding  a hydrostatic layer computed using  
the MLT to model A. This increased  the 
depth from 2.1 Mm to 2.8 Mm. As the 
turbulent pressure near the bottom of model A was small 
(about 1 \% of the gas pressure), in this 
case we did not need to relax the hydrostatic lower layer separately. 
\subsubsection{Effect on kinetic energy and turbulent pressure}
To demonstrate  the effect 
of the lower boundary on the turbulent flow, we computed  the  ratio of 
the horizontal kinetic energy to the gas pressure.  This  is equivalent to the 
horizontal turbulent Mach number squared, i.e. 
\be
\chi_{\rm horiz} = \frac{1}{2}\overline{\rho}({v_x''}^2+{v_y''}^2)/\overline{P}=\frac{1}{2}({v_x''}^2+{v_y''}^2)/{c_s}^2
\ee
Fig. \ref{kehoriz}  shows  $\chi_{\rm horiz}$
for models KC2, A and B. In KC2 the magnitude of
$\chi_{\rm horiz}$ shoots up at the base.
The net effect of the fast downflows striking the lower boundary  was to
speed up the overall flow.
This increased the total turbulent kinetic energy $\chi$ throughout the box (Fig. \ref{ke}).  
 
The upturn near the bottom of each layer,
is  clearly reduced as the boundary is moved deeper, and 
is almost eliminated when the depth is 2.8 Mm.
This  problem is much less severe at the top boundary
because the underlying region is sub-adiabatic (the radiative zone).  The vertical velocity  is 
already small when the flow hits the top boundary. 
The effect on $P_{\rm turb}$ of moving down the lower boundary is shown in Fig. \ref{ppturb}.
In  model KC2, $P_{\rm turb}$ drops sharply as the impenetrable  bottom is approached.
While for  A and B it has a much smoother decay.

\subsubsection{Effect on superadiabaticity}  
The run of
$\na -\na_{\rm ad}$ for the MLT,
and models KC2,  and C was  shown previously in Fig. \ref{salr}.
When  the SAL is moved outwards by turbulent pressure, 
the convective efficiency is reduced and radiation is forced to carry more
of the total flux. This should result in an increased  height of
the SAL   compared to the MLT.
However, in KC2 the convection is overly turbulent 
and  is thus (incorrectly) more efficient than convection computed 
via  the MLT.
This results in a  drop in the height of the SAL
compared to the MLT.

\subsection{The influence of the upper impenetrable boundary}
\label{upperbc}
As the top boundary is also impenetrable we need to ensure 
that its position does not 
effect the convection either. We need to prove that the decrease of vertical velocity
between  the convection surface (where the depth = 0 Mm) and  the top of the box, is
not due to the top boundary.  Rather it should be due to  the stable layer
at the top, i.e. convection-radiation losses.

To address this issue we damped the horizontal velocity at the top 
by replacing the stress free  boundary
with a no-slip
top. The flow is then relaxed and statistics are gathered as usual.
Fig. \ref{hbdry} shows $\na-\na_{ad}, v_x'' $ and $v_z''$  
for simulations with  slip and no-slip top boundaries.
The velocities are non-dimensionalised  by the local isothermal sound speed. This conveniently enables 
them to be on the same axis as  $\na-\na_{ad}$. 
Apart from the top 100 km, where  the two $v_x''$'s  diverge 
because of the different top boundary conditions, the rms vertical and horizontal velocities 
are nearly the same for the no-slip
and stress free top boundaries.
This implies the drop in  vertical velocity near the top is primarily due to
to convection to radiation losses, rather than the top boundary.
Furthermore, the horizontal velocity in the upper atmosphere
does not affect the SAL structure (i.e. $\na-\na_{ad}$) much.

\subsection{The influence of the domain width}
Models B, C,  D and E have widths of 1.35 Mm,  2.7 Mm, 2.7 Mm  and 5.4 Mm.  
To judge the effect of width on simulations without changing the grid spacing, we should  compare 
B to C and D to E. 
All four models have similar $P_{\rm turb}$ and $\na - \na_{\rm ad}$  
(Figs. \ref{pturbwidth} and \ref{salwidth}). 
Increasing the width  seems to have a minor  
effect on either $P_{\rm turb}$ or $\na - \na_{\rm ad}$.
In general  it is essential to resolve turbulent motions inside the SAL region.
As this is only about 250 km thick, the grid spacing in the SAL need to be very small. 
To show that all our simulations have resolved the SAL, we 
have plotted the individual vertical grid points on the figure.

The variation of $\chi$ with domain width is more interesting. 
This is shown in  Fig. \ref{kewidth}.
When the width is increased from  1.35 Mm  to 2.7 Mm, 
$\chi$ increases (especially near the top). 
However, when the width is increased from 2.7 Mm to 5.4 Mm 
there is only a very small change in $\chi$.
A box width of 2.7 Mm seems to be a sufficient 
to resolve $\chi$ for  solar granules.

To provide further evidence that 2.7 Mm is a large enough box width,
we computed $v_x'', v_y''$ and $v_z''$ for  
models D and E. Fig. \ref{min_width} shows all three velocity components for both D and E. 
Doubling the width has only a minor effect on 
any particular velocity component. The small differences in the deeper 
part,   
are due to insufficient convergence. As the domain has 
periodic lateral boundaries, eventually the horizontal velocities 
should
all be the same.

Fig. \ref{granulewidth} shows contour plots of the instantaneous vertical velocity, 
for a horizontal cross-section, for models of different widths.
The uppermost frame shows that not even a single granule can fit in the box 
when the width is 1.35 Mm, while about 2 fit easily in to the 
box in the second frame. 
In the final frame about 9 or 10 larger cells 
fit in to the box. Due to computational restrictions the last figure 
is computed on a coarser mesh than the previous three frames so  the granules are not 
clearly depicted. 

\subsection{The influence of numerical resolution}
While the main point of these tests is to show that the walls of the computational box
can have a significant effect on granular convection, we 
can also partially address  the effect of changing the horizontal  
grid spacing. Models C and D  produced very similar  $P_{\rm turb}$ and 
$\na -\na_{\rm ad}$, 
while $\chi$ differed only slightly. As $\chi$ depends on the horizontal as well as the vertical  velocity, 
this probably reflects the 
sensitivity of the horizontal component of kinetic energy to horizontal grid resolution. 
This is particularly noticeable near the top where the flow is mostly horizontal.


\newpage
\vspace{3mm}
\begin{figure}
\epsfig{file=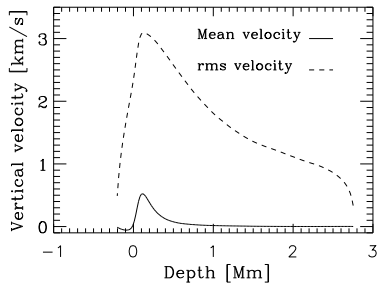,width=6.cm}
\caption{\label{vcomparison} The mean and rms vertical velocity  in model C.} 
\end{figure}
\begin{figure}
\epsfig{file=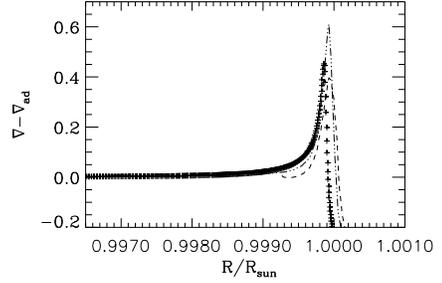,width=6.cm}
\caption{\label{salr}Superadiabaticity versus fractional radius.
The crosses are from the 1D stellar model (MLT), the dashes are for model KC2 and the
triple dot-dash is for model C (see appendix for details). 
In both KC2 and  C the original (MLT) convective boundary is moved out by turbulent pressure.
}
\vspace{3mm}
\end{figure}   
\begin{figure}
\epsfig{file=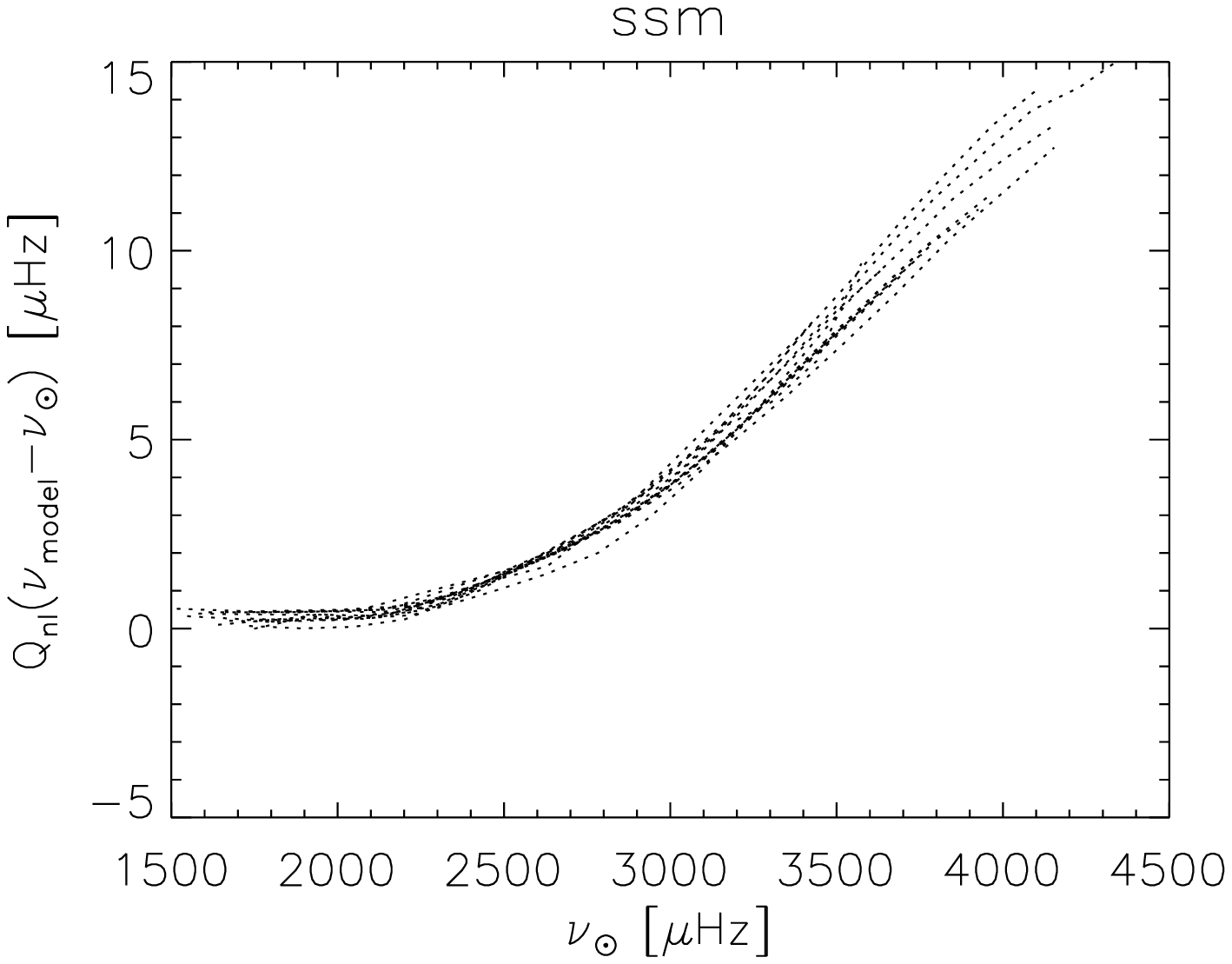,width=6.cm}
\caption{\label{pmodessm}   The difference between the observed and the computed {\it p}-mode
frequencies for a Standard Solar model, SSM (see text).}  
\epsfig{file=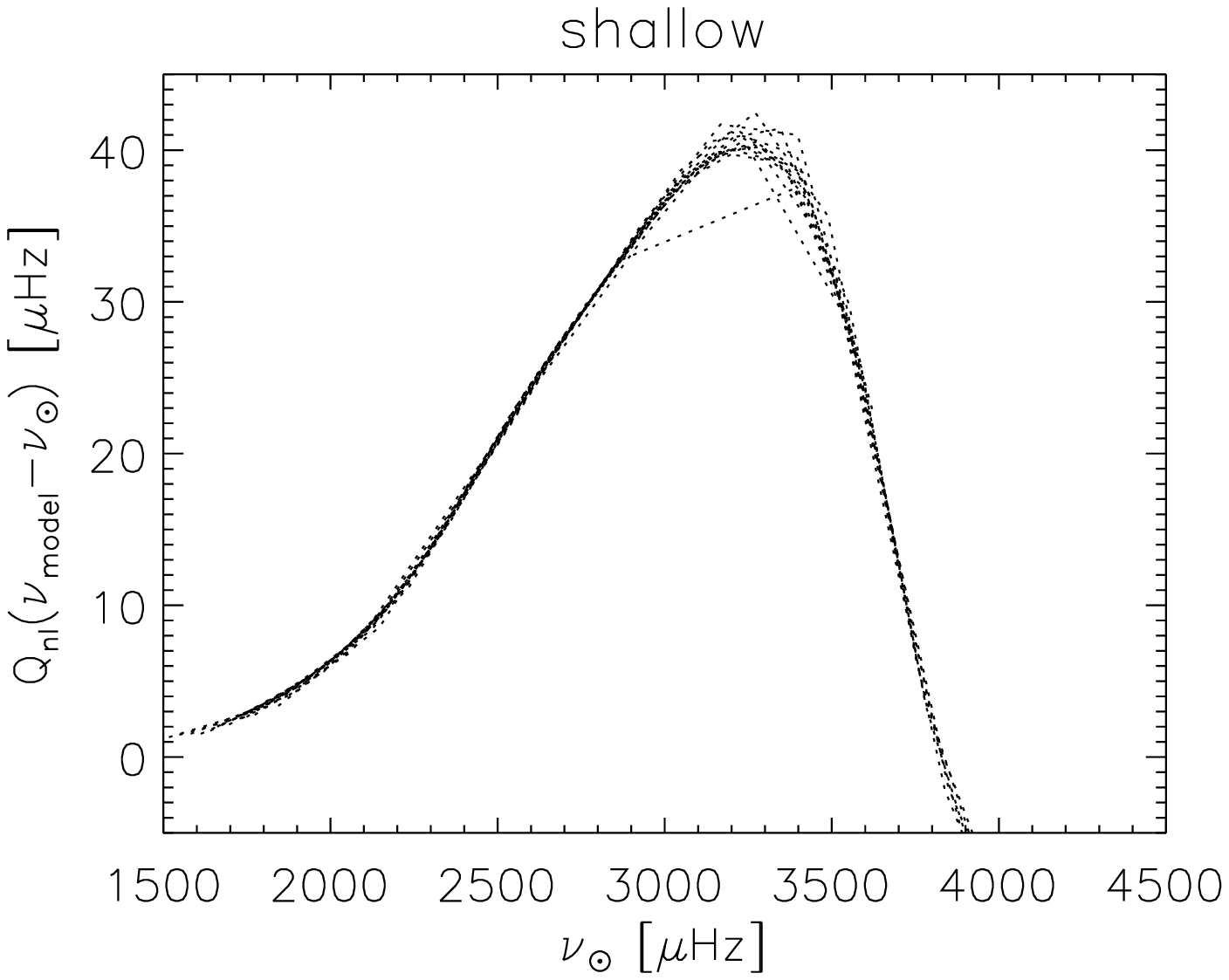,width=6.cm}
\caption{\label{pmodeshallow}The difference between the observed and the computed {\it p}-mode
frequencies for a solar model with turbulence included from model KC2 (see appendix).} 
\epsfig{file=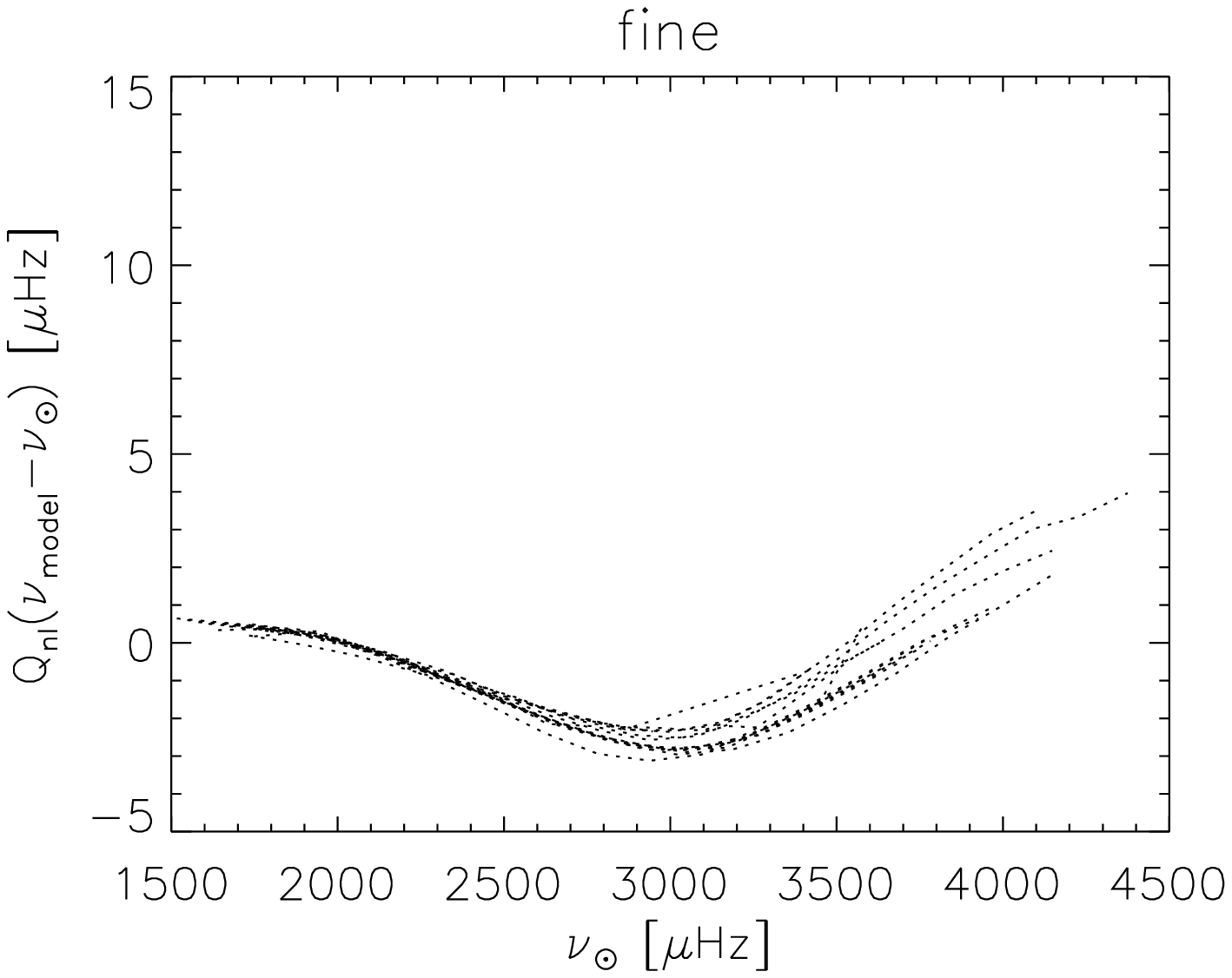,width=6.cm}
\caption{\label{pmodedeep} The difference between the observed and the computed {\it p}-mode
frequencies with turbulence included from model C (see appendix).}
\end{figure}
\newpage
\begin{figure}
\epsfig{file=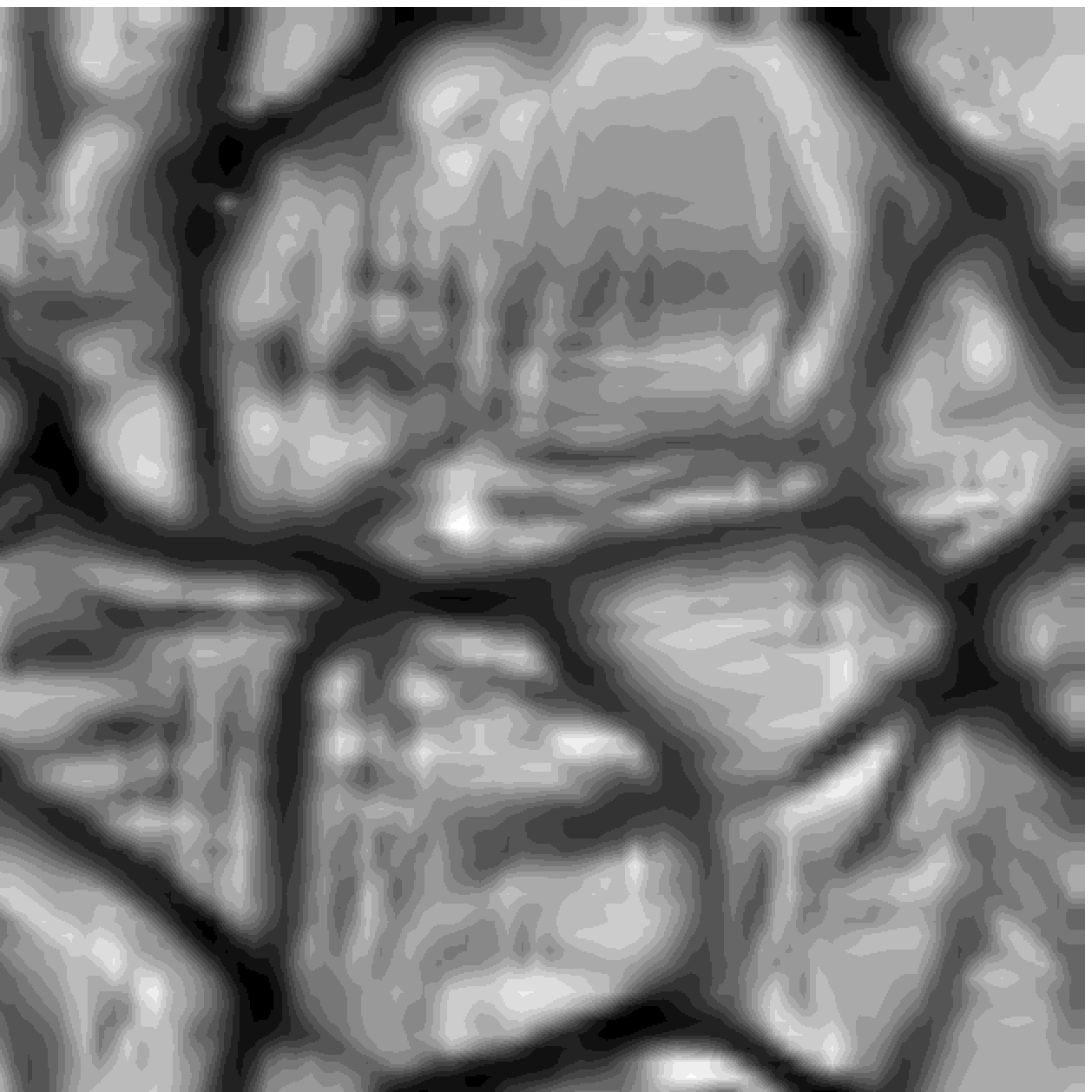,height=2in}
\epsfig{file=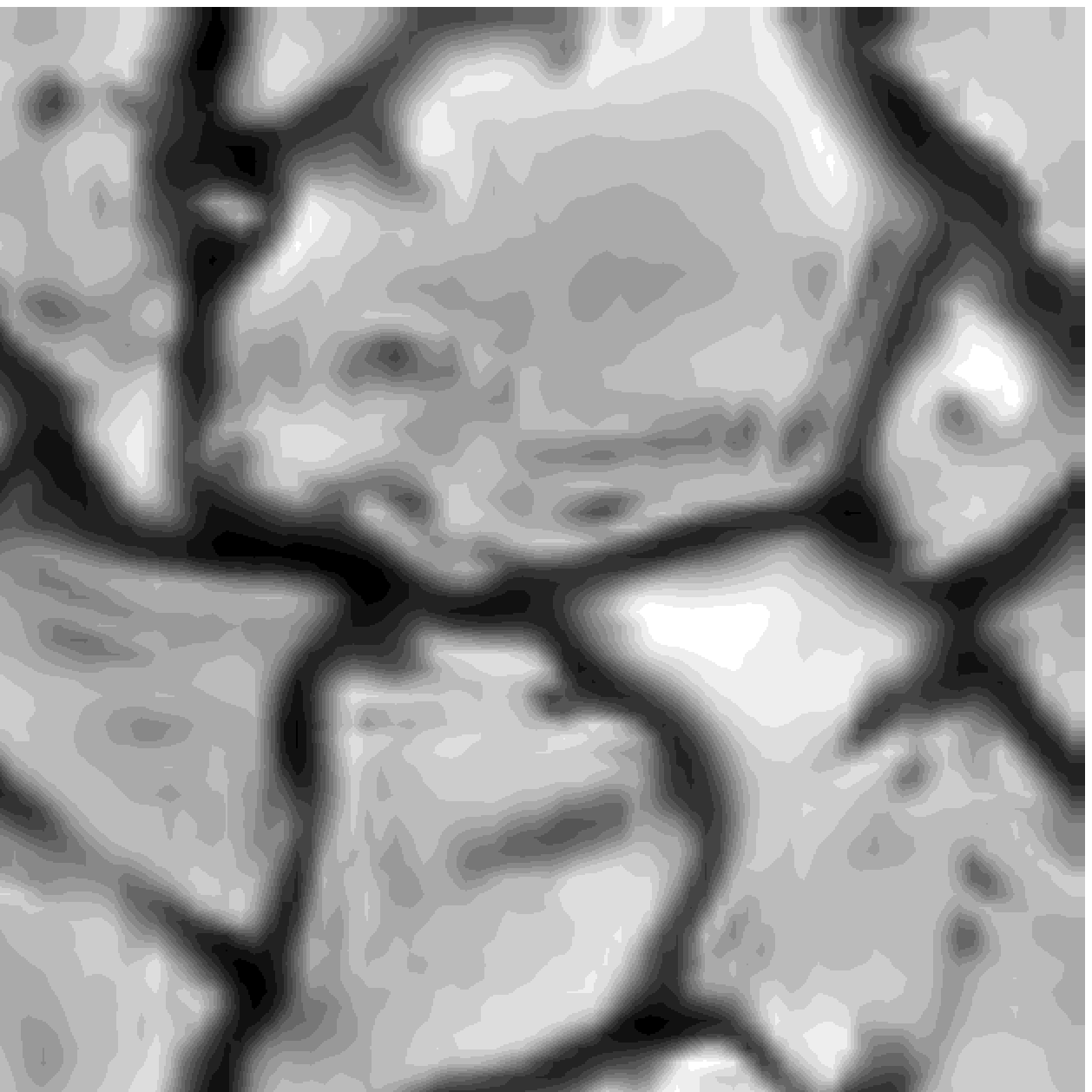,height=2in}
\epsfig{file=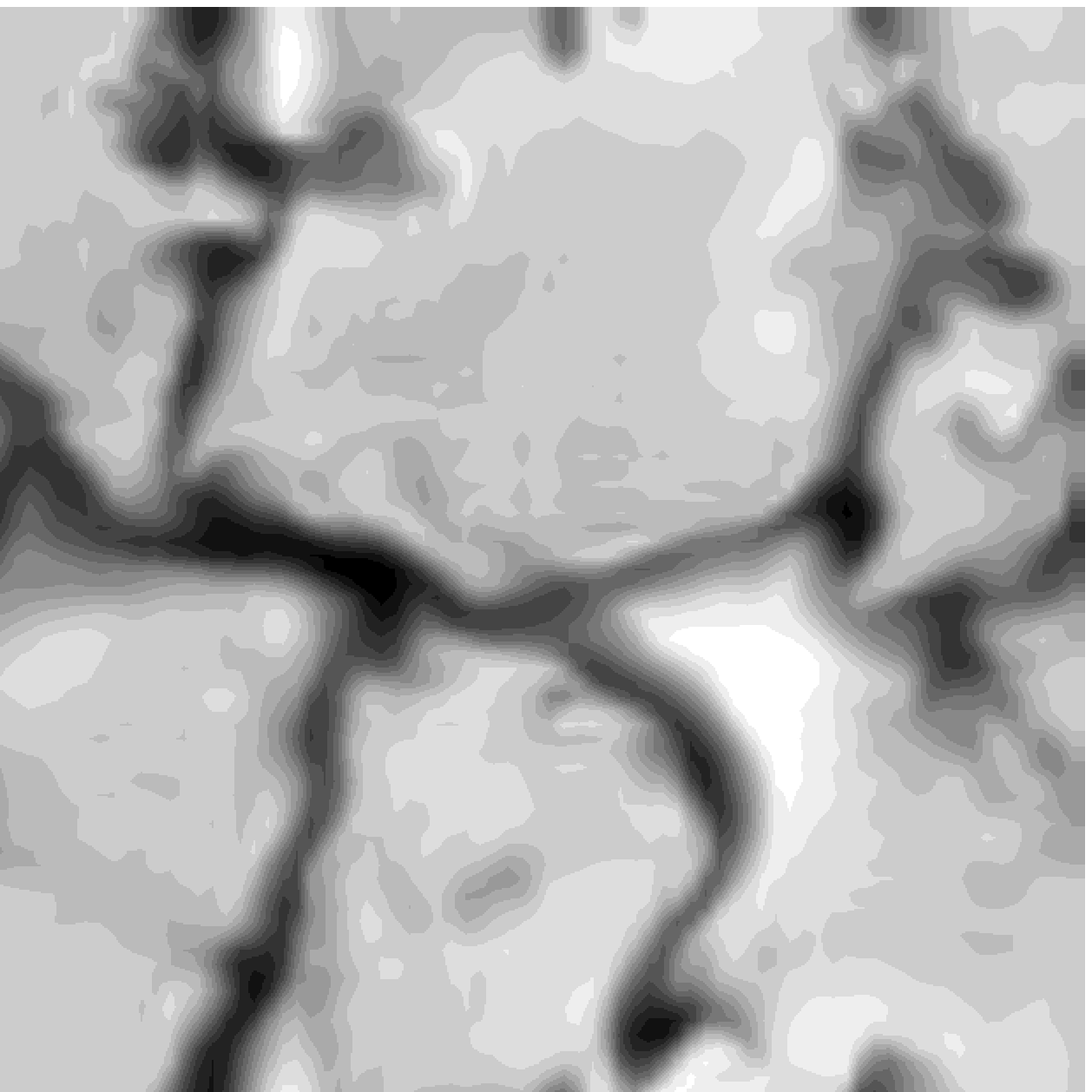,height=2in}
\epsfig{file=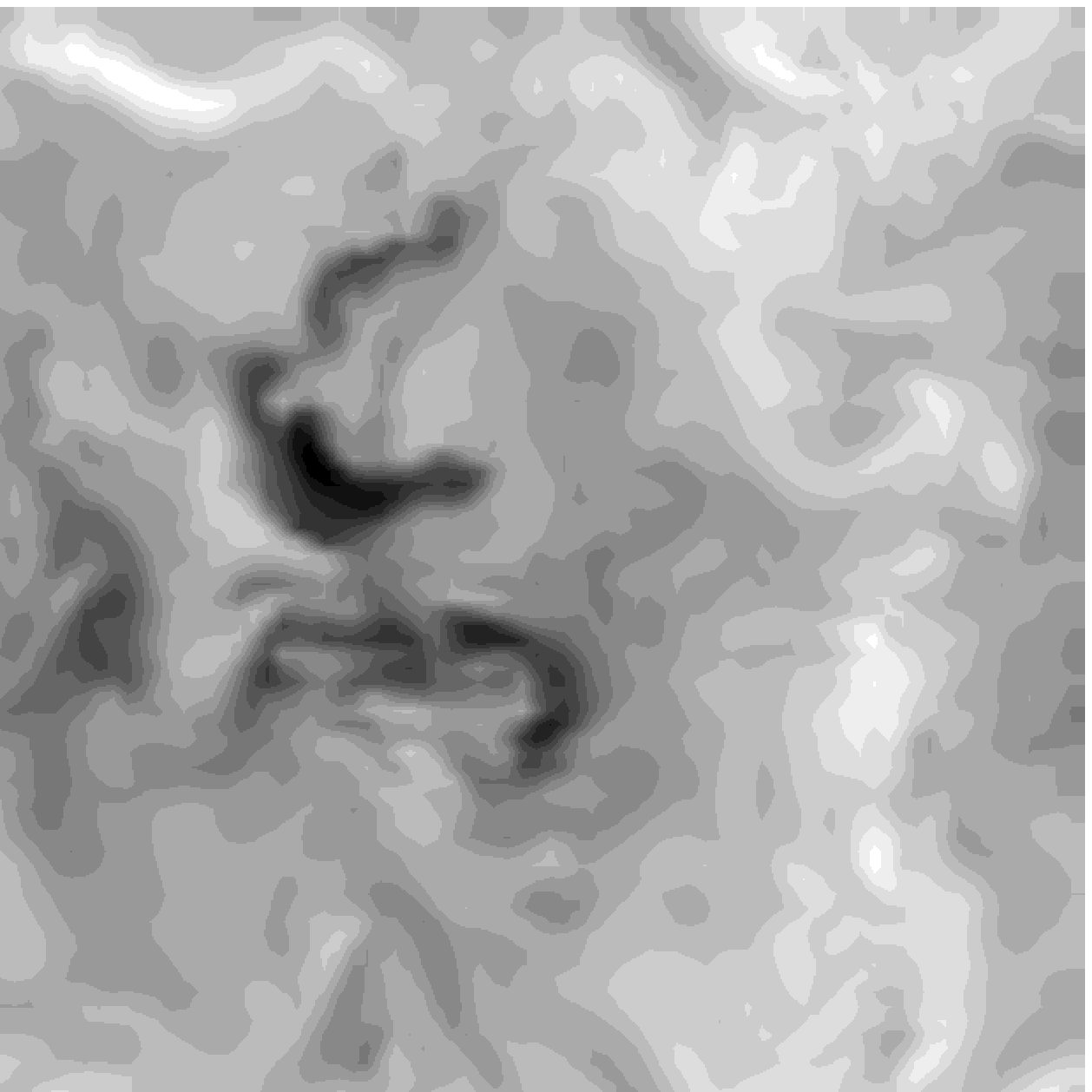,height=2in}
\caption{\label{granr}Contours of vertical velocity at one instant in time. The darker regions are downflows,  
and the lighter regions  are upflows. 
From top to bottom, the frames are at depths of 
-0.14 Mm, 0.03 Mm, 0.2 Mm and 1.0 Mm. Positive  depths are measured  inwards from  the solar surface. 
The width of each frame  is 3.75 Mm.
The small
parallel vertical lines in the first frame indicate 
significant grid oscillations. By slightly increasing the SGS viscosity 
at the top, the oscillations  are damped out before starting  the actual statistical computations. 
The contours themselves suggest that the 
granules remain correlated throughout the SAL 
(which is between 0 Mm and about 0.25 Mm). By  1 Mm there is no sign of granulation.}
\end{figure}
\newpage
\begin{figure}
\epsfig{file=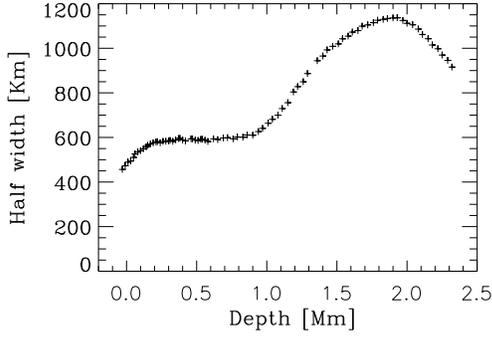,height=2in}
\caption{\label{hw} The half width of the 2-point vertical correlation 
length of vertical velocity at different depths. The right side of  the plateau is  3 PSH 
above the bottom of the box.
Between the surface and a depth of 1 Mm, the half-width is nearly constant.}   
\end{figure}
\begin{figure}
\epsfig{file=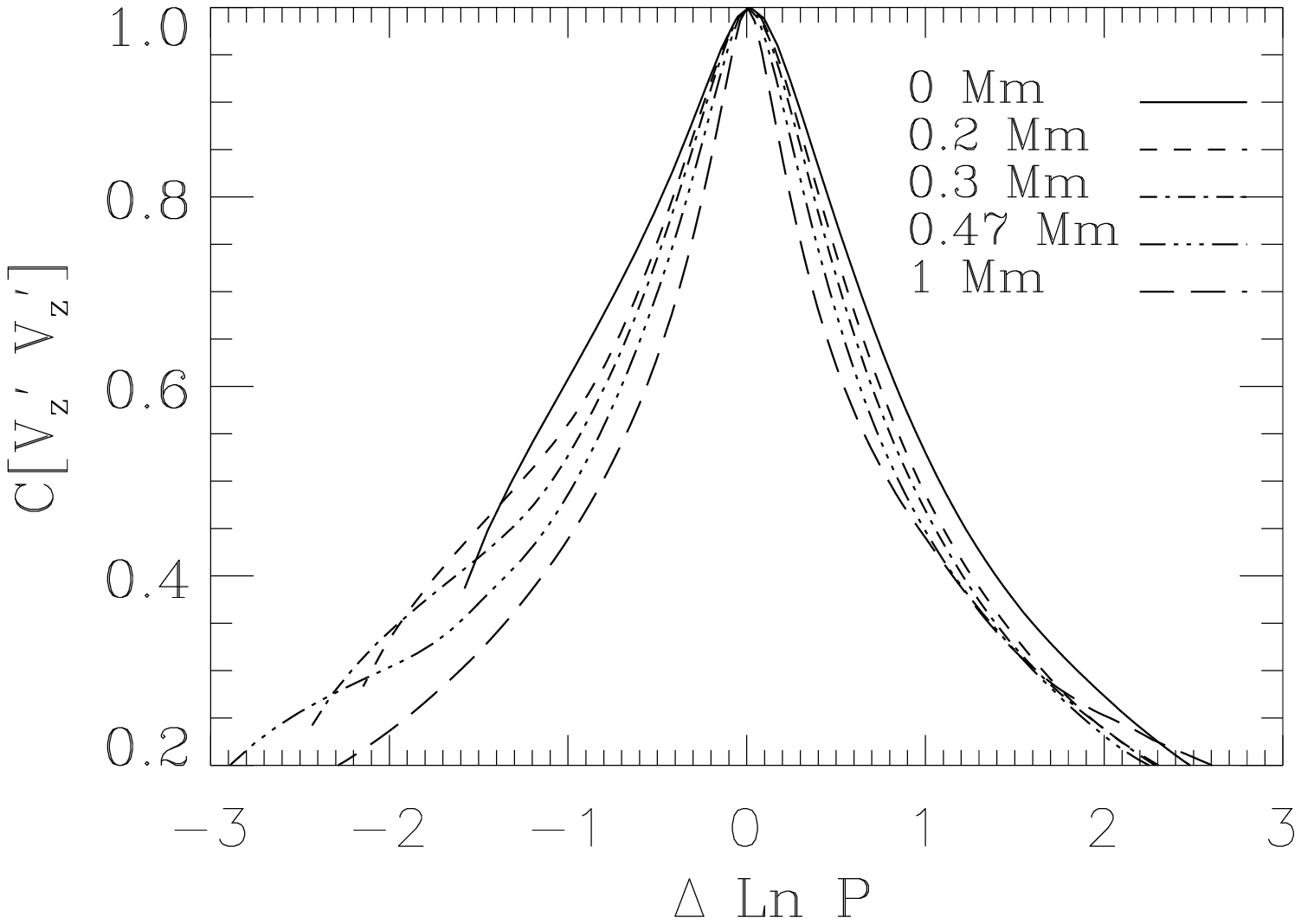,height=2in}
\caption{\label{vzvz} 2-point vertical velocity correlation 
at 5 different depths. Unlike the case of deep 
convection, the velocity correlation does not scale with pressure scale height } 
\epsfig{file=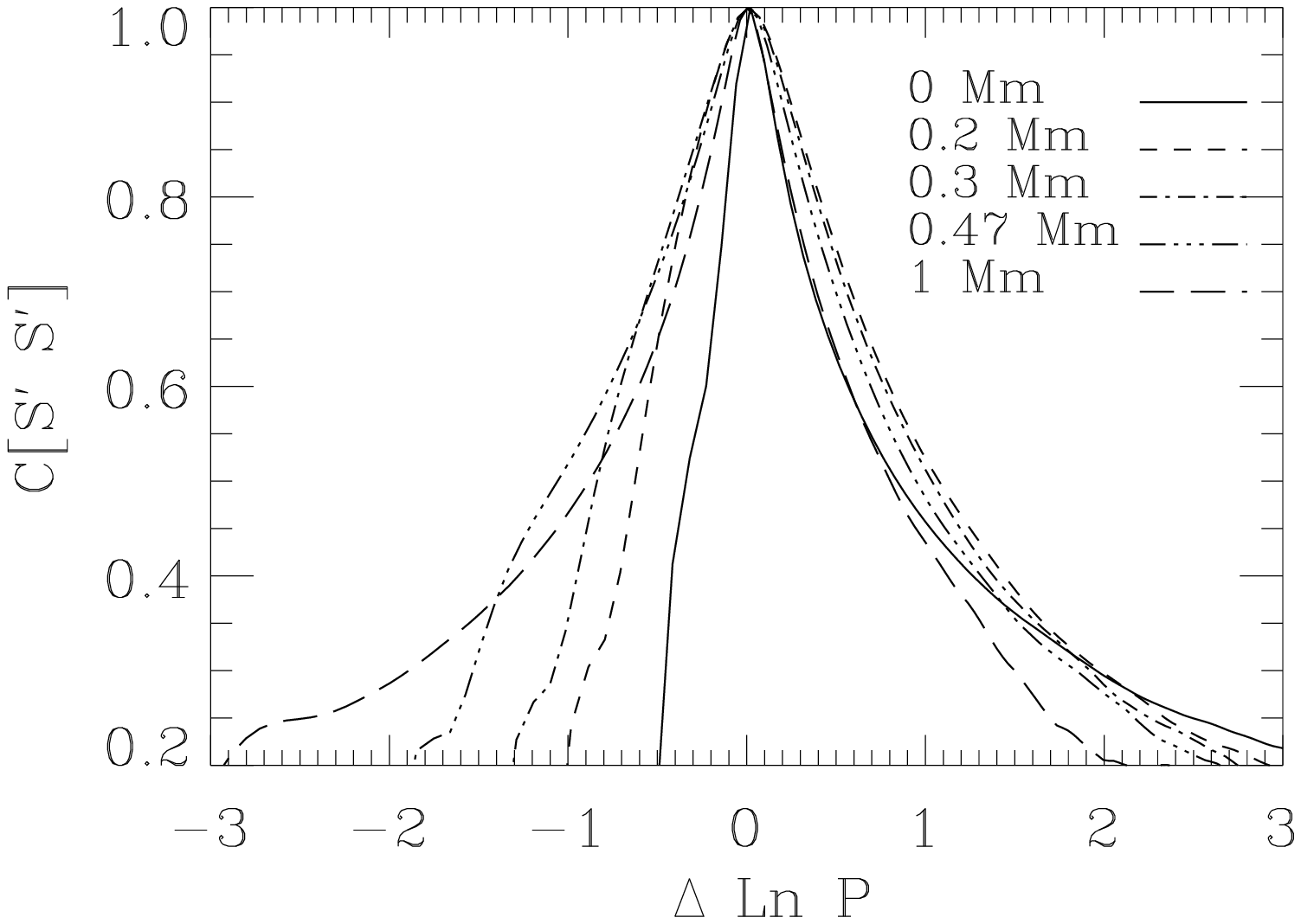,height=2in}
\caption{\label{ss} 2-point entropy  correlation
at the same levels as in the previous figure. Close to the surface, ascending fluid parcels loose their 
entropy identity much sooner than descending parcels.}
\epsfig{file=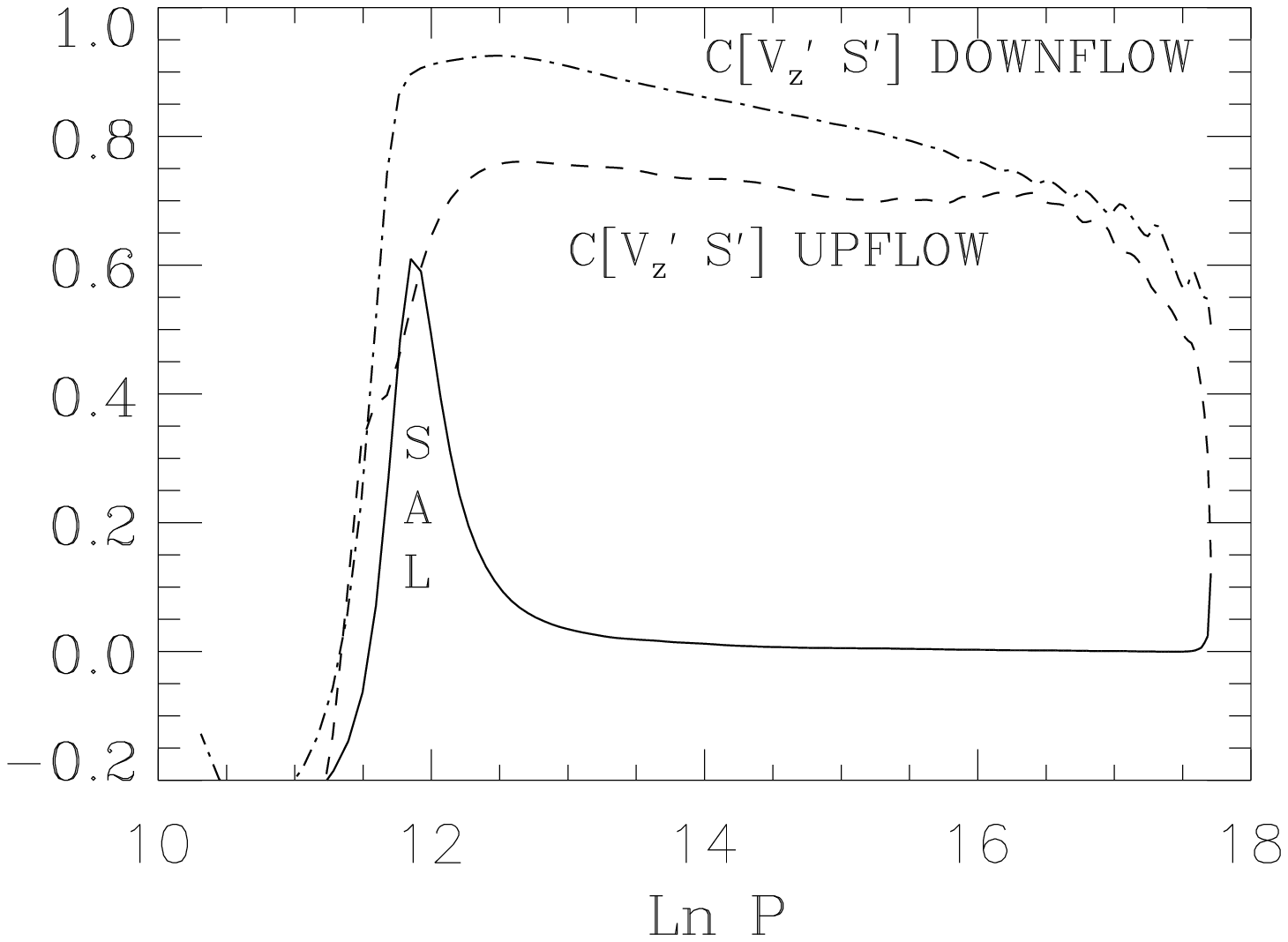,height=2in}
\caption{\label{cvzs} Correlation coefficient between
entropy and vertical velocity for upflows and downflows.
Also plotted is $\na - \na_{\rm ad}$ for comparison.}
\end{figure}        
\begin{figure}
\epsfig{file=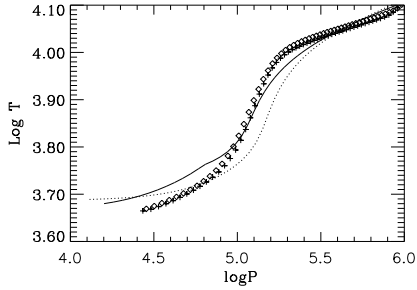,width=6.cm}
\caption{\label{f4}Hydrostatic structure  of two 1D stellar models (see text) for KC (dotted) and KC2  (solid).
The mean  thermal structure from the  hydrodynamic  simulations
is also shown for  KC (diamonds) and KC2 (crosses).  Despite different
initial
stellar models, the turbulence causes both models to eventually converge to the same equilibrium state.}
\end{figure}
\begin{figure}  
\epsfig{file=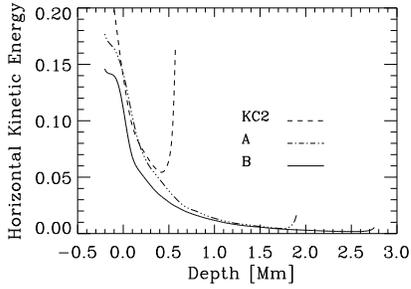,width=6.cm}
\caption{\label{kehoriz}Non-dimensional horizontal turbulent kinetic energy per unit mass,
for simulations of different depths. Note
the speeding up of the flow at the lower stress free boundary in KC2.} 
\end{figure}
\begin{figure}
\epsfig{file=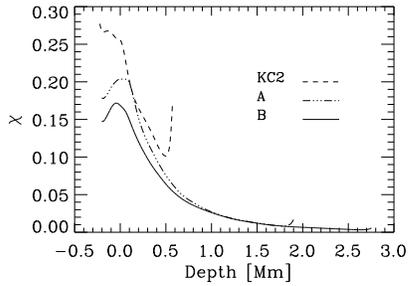,width=6.cm}
\caption{\label{ke} Non-dimensional turbulent kinetic energy per unit mass, $\chi$
for simulations of different depths.}
\vspace{3mm}
\end{figure}
\begin{figure}
\epsfig{file=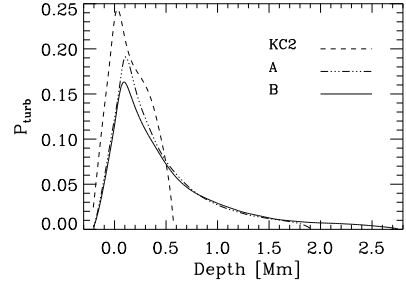,width=6.cm}
\caption{\label{ppturb}Ratio of turbulent pressure  to gas pressure, in this case denoted by $P_{\rm turb}$,
for simulations of different depths.} 
\end{figure}

\vspace{3mm}
\begin{figure}
\epsfig{file=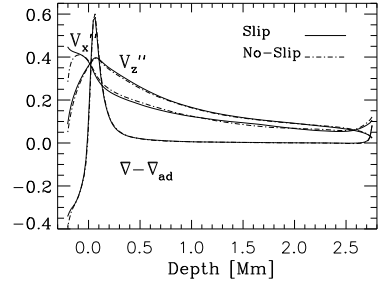,width=6.cm}
\caption{\label{hbdry}
Model D with a slip (stress free) and a no-slip top boundary.
The figure includes horizontal and vertical turbulent velocities and superadiabaticity.
The velocities have been  scaled by the local sound speed.} 
\end{figure}

\begin{figure}
\epsfig{file=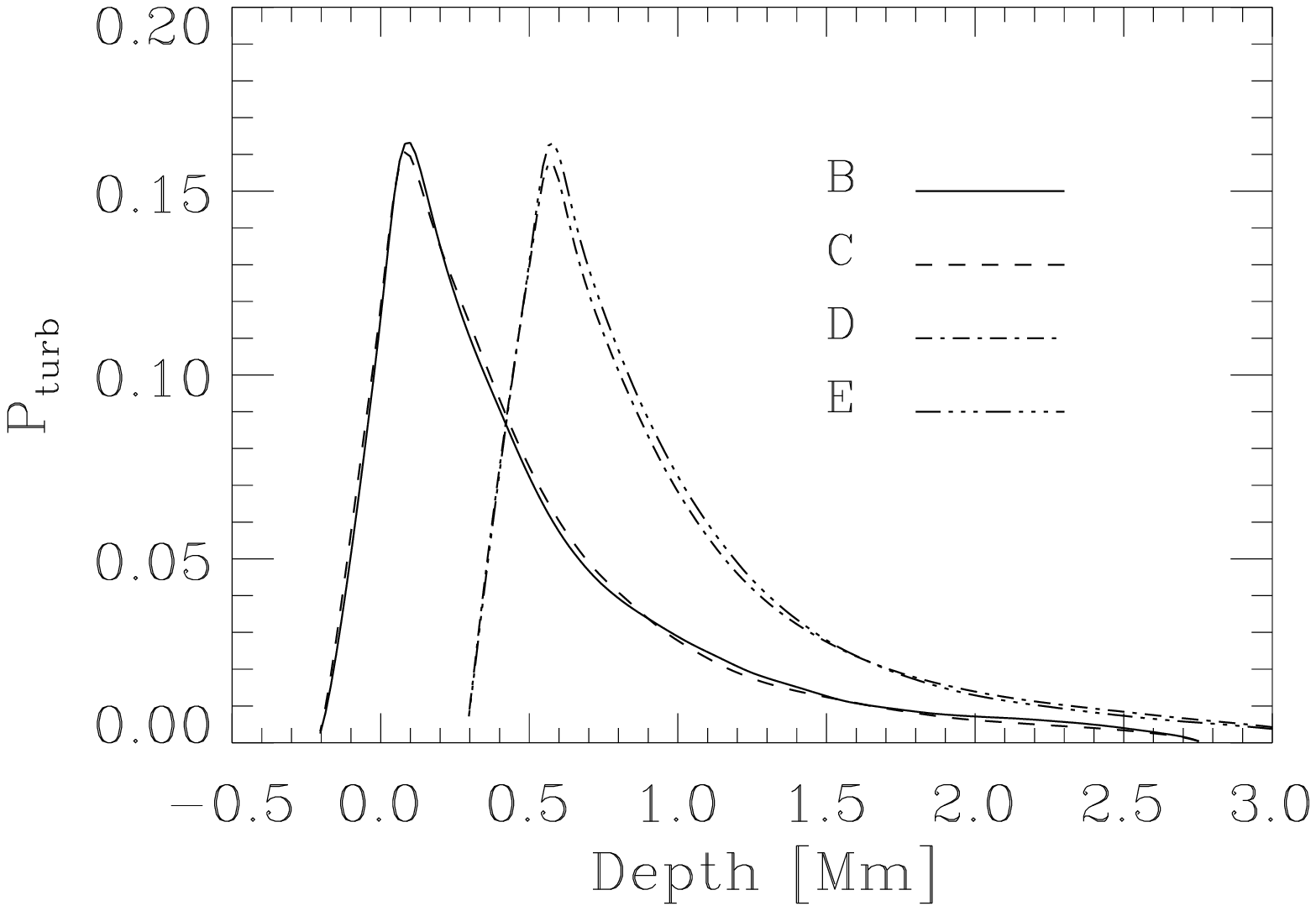,width=6.cm}
\caption{\label{pturbwidth} 
Turbulent pressure as a function  of domain width. $P_{\rm turb}$  is the ratio of turbulent pressure to 
gas pressure. 
The plot for D and E has been shifted by 0.5 Mm for clarity. Without the shift the peaks would coincide.}
\vspace{3mm}
\epsfig{file=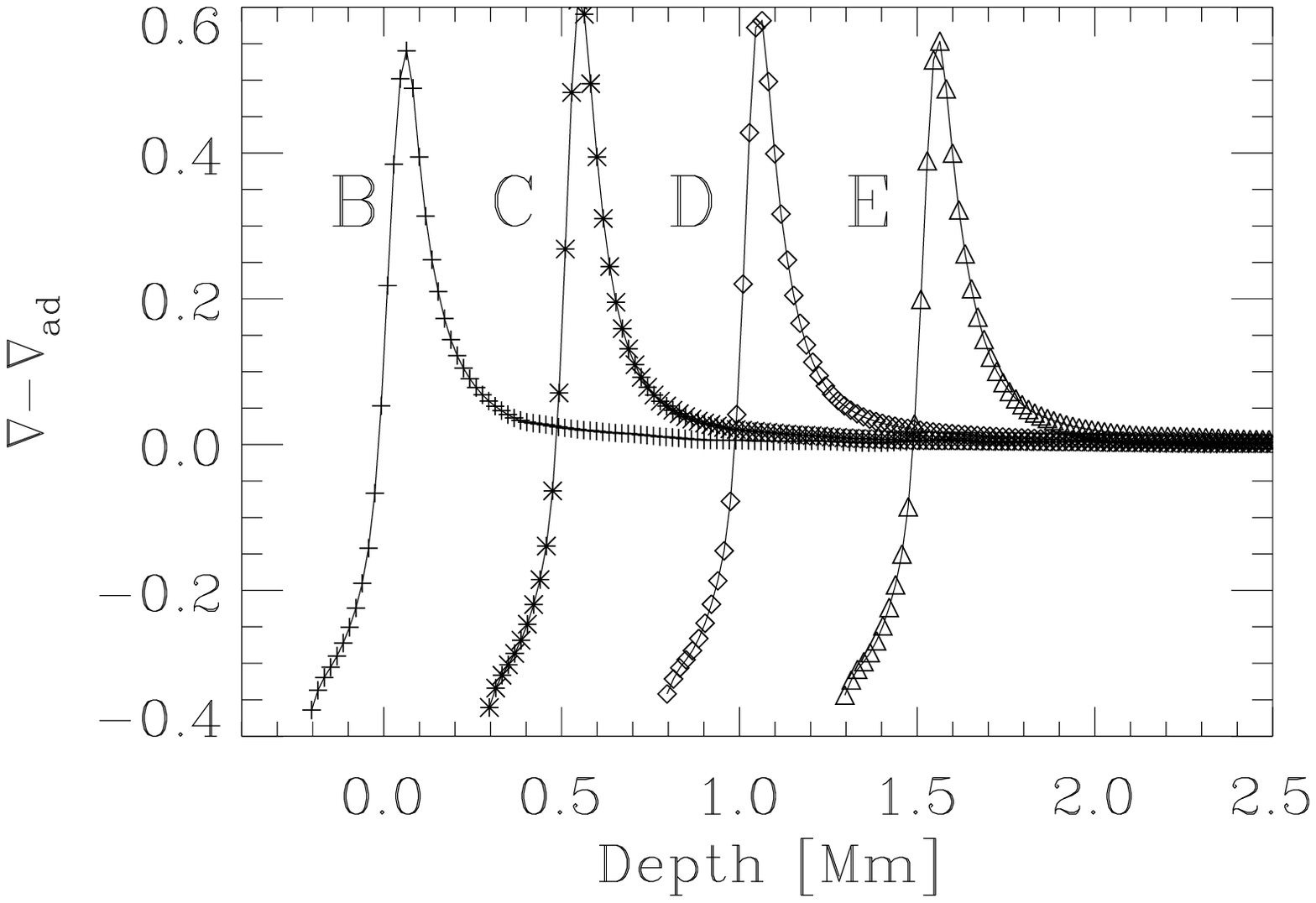,width=6.cm}
\caption{\label{salwidth}
Superadiabaticity  as a function  of domain width for models B, C, D and E.
The vetical grid points are individually marked to show that the SAL is well resolved.
The plots have been spaced  by 0.5 for clarity. Without the horizontal spacing  all the plots would have 
zero superadiabaticity at a depth of 0 Mm.}
\end{figure}
\begin{figure}
\epsfig{file=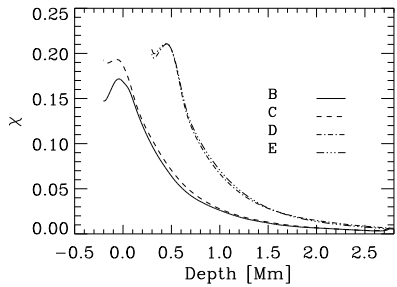,width=6.cm}
\caption{\label{kewidth}
Non-dimensional turbulent kinetic energy  for different domain widths. 
The plot for D and E has been shifted by 0.5 Mm for clarity. }
\end{figure}
\begin{figure}
\epsfig{file=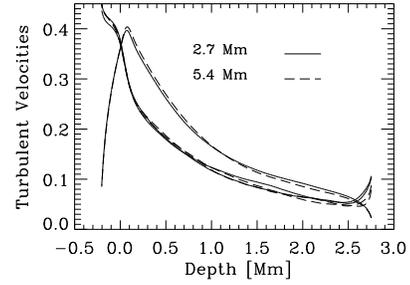,width=6.cm}
\caption{\label{min_width}
The three components of turbulent velocity (scaled by the isothermal sound speed), for simulations D and E.
As the change in width has little effect on any component of the velocity, 2.7 Mm seems to
be a sufficient box width.}
\end{figure}

\newpage
\begin{figure}
\epsfig{file=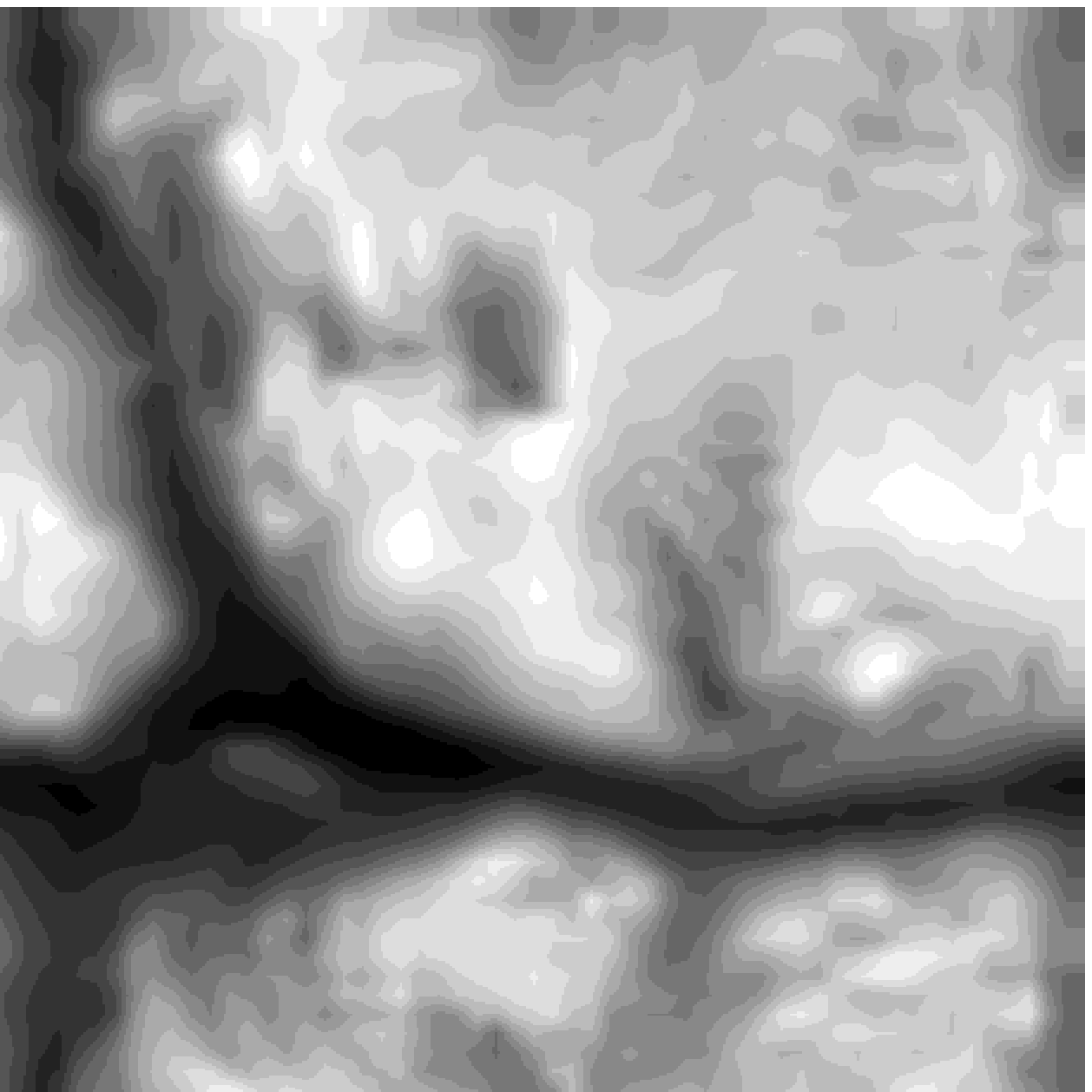,height=2in}
\epsfig{file=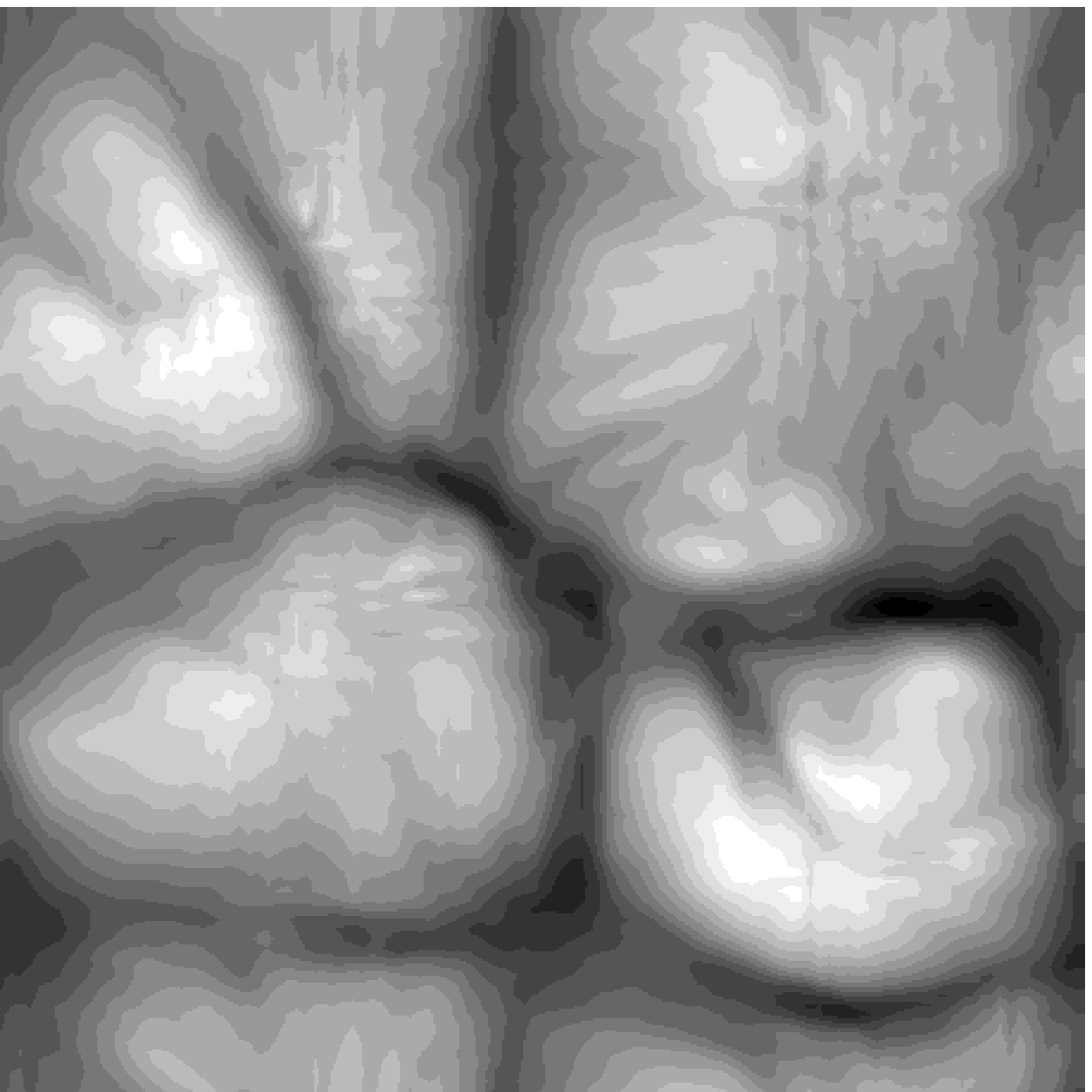,height=2in}
\epsfig{file=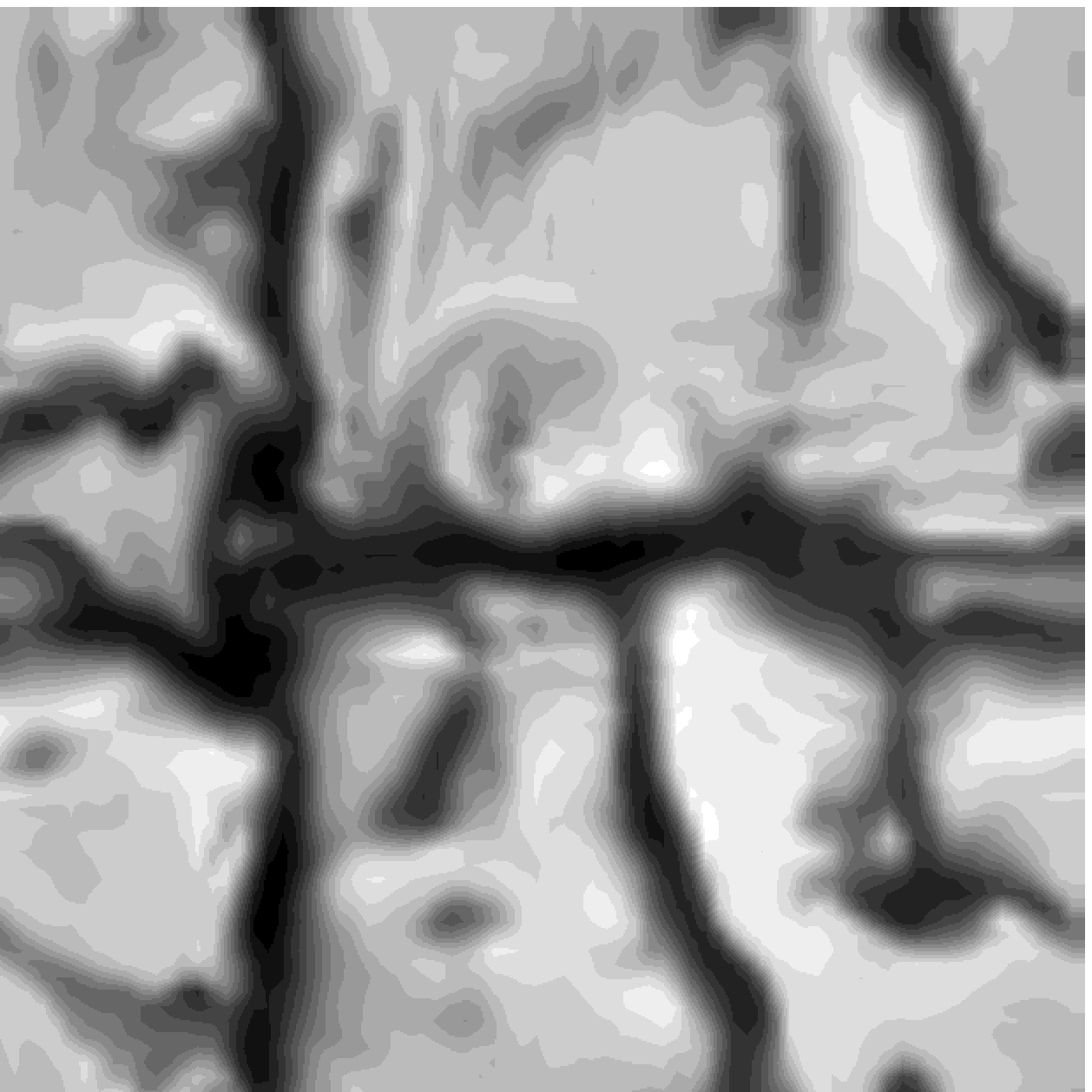,height=2in}
\epsfig{file=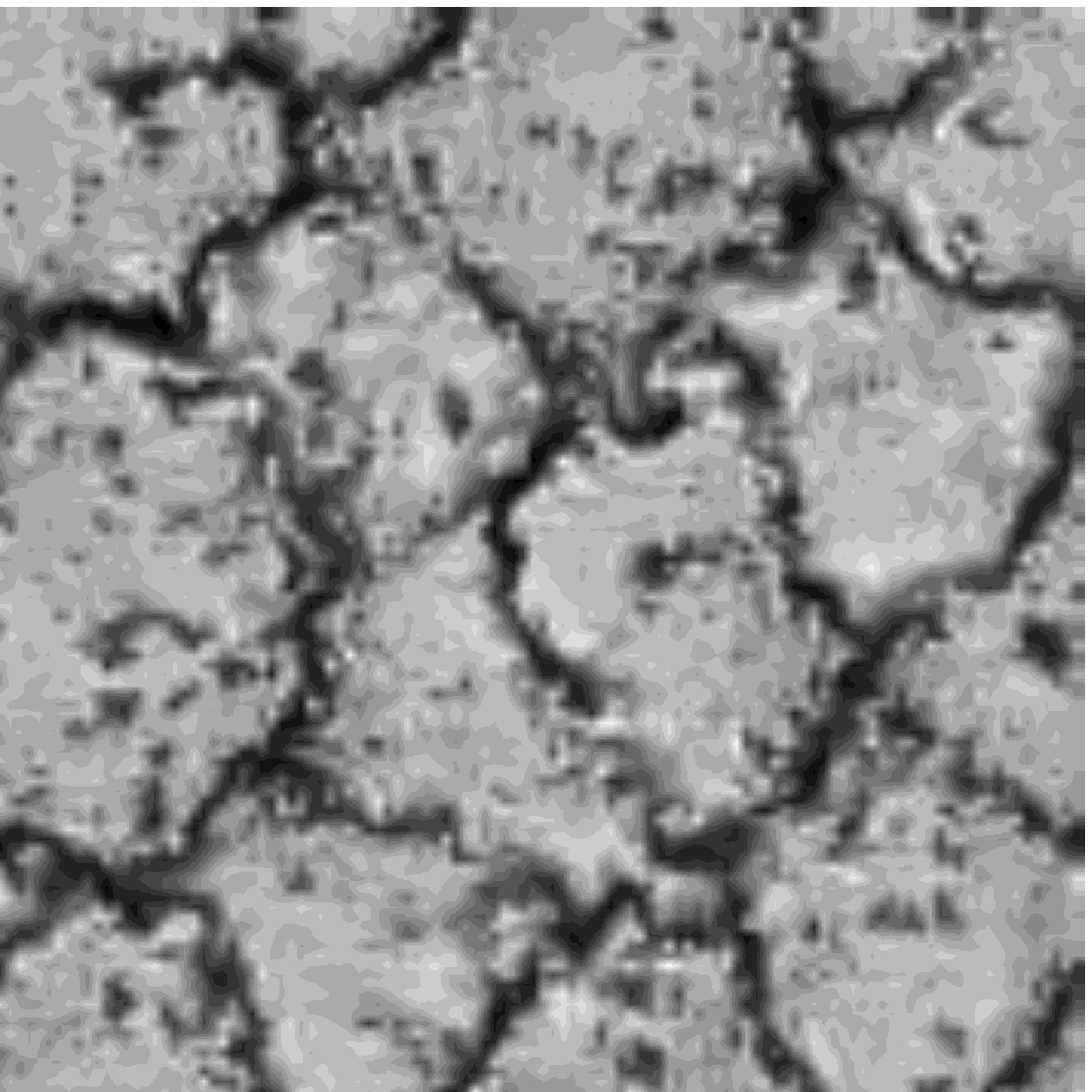,height=2in}
\caption{\label{granulewidth}Snapshot of vertical velocity contours 
for different domain widths.
From  top to bottom
the domain widths are  1.35, 2.7, 3.75 and 5.4 Mm, respectively. Due to computational restrictions the 
last frame has half the resolution than the other 3.} 
\end{figure}   
\end{document}